\documentclass[12pt]{spieman}  
\usepackage{amsmath,amsfonts,amssymb}
\usepackage{graphicx}
\usepackage{setspace}
\usepackage{tocloft}
\usepackage{lineno}
\usepackage{longtable}

\newcommand{\rev}[1]{#1} 

\usepackage{hyperref}
\usepackage{xcolor}

\title{Lessons learned from SPHERE for the astrometric strategy of the next generation of exoplanet imaging instruments}

\author[a,b,*]{Anne-Lise Maire}
\author[c,d]{Maud Langlois}
\author[e]{Philippe Delorme}
\author[e,f]{Ga\"el Chauvin}
\author[g]{Raffaele Gratton}
\author[d]{Arthur Vigan} 
\author[h]{Julien H. Girard}
\author[i]{Zahed Wahhaj}
\author[b]{J\"org-Uwe Pott} 
\author[j]{Leonard Burtscher} 
\author[k]{Anthony Boccaletti}
\author[e]{Alexis Carlotti}
\author[b]{Thomas Henning}
\author[j]{Matthew A. Kenworthy} 
\author[k]{Pierre Kervella}
\author[h]{Emily L. Rickman}
\author[k,l]{Tobias O. B. Schmidt}
\affil[a]{Universit\'e de Li\`ege, STAR Institute, Li\`ege, Belgium}
\affil[b]{Max-Planck-Institut f\"ur Astronomie, Heidelberg, Germany}
\affil[c]{Centre de Recherche Astrophysique de Lyon, CNRS, Universit\'e Lyon 1, ENS Lyon, Saint-Genis-Laval, France}
\affil[d]{Aix-Marseille Universit\'e, CNRS, CNES, Laboratoire d'Astrophysique de Marseille, Marseille, France}
\affil[e]{CNRS, IPAG, Universit\'e Grenoble Alpes, Grenoble, France}
\affil[f]{CNRS/INSU, Universidad de Chile, Departamento de Astronom\'ia, Unidad Mixta Internacional Franco-Chilena de Astronom\'ia, Santiago, Chile}
\affil[g]{INAF – Osservatorio Astronomico di Padova, Padova, Italy}
\affil[h]{Space Telescope Science Institute, Baltimore, Maryland, United States}
\affil[i]{European Southern Observatory, Vitacura, Chile}
\affil[j]{Leiden Observatory, Leiden University, Leiden, The Netherlands}
\affil[k]{Universit\'e PSL, CNRS, Sorbonne Universit\'e, Universit\'e de Paris, LESIA, Observatoire de Paris, Meudon, France}
\affil[l]{Hamburger Sternwarte, Hamburg, Germany}

\cftpagenumbersoff{figure}
\cftpagenumbersoff{table} 
\begin{document} 
\maketitle

\begin{abstract}
Measuring the orbits of directly-imaged exoplanets requires precise astrometry at the milliarcsec level over long periods of time due to their wide separation to the stars ($\gtrsim$10~au) and long orbital period ($\gtrsim$20~yr). To reach this challenging goal, a specific strategy was implemented for the instrument Spectro-Polarimetric High-contrast Exoplanet REsearch (SPHERE), the first dedicated exoplanet imaging instrument at the Very Large Telescope of the European Southern Observatory (ESO). A key part of this strategy relies on the astrometric stability of the instrument over time. We monitored for five years the evolution of the optical distortion, pixel scale, and orientation to the True North of SPHERE images using the near-infrared instrument IRDIS. We show that the instrument calibration achieves a positional stability of $\sim$1~mas over 2$''$ field of views. We also discuss the SPHERE astrometric strategy, issues encountered in the course of the on-sky operations, and lessons learned for the next generation of exoplanet imaging instruments on the Extremely Large Telescope being built by ESO.
\end{abstract}


{\noindent \footnotesize\textbf{*}Anne-Lise Maire,  \linkable{almaire@uliege.be} }

\begin{spacing}{1}   

\section{Introduction}

Orbital monitoring of exoplanetary and stellar systems is fundamental for analyzing their architecture, their dynamical stability and evolution, and even tracing back their mechanisms of formation. In the context of imaging surveys for giant exoplanets\cite{Biller2013, Galicher2016, Langlois2021}, high-precision relative astrometry is required. Relative astrometry is instrumental for determining the nature of the faint sources detected near the targeted stars. The fields of view (FoVs) used in direct imaging are typically too small for absolute astrometry so that astrometry relative to the targeted star is used instead. Multiple-epoch monitoring enables to test if the candidate companions are comoving with similar proper and parallactic motion than the host star by rejecting contamination by stationary (or slowly moving with the local field) background or foreground source. More precise measurements allow for faster confirmations, which is critical in a context of the strong international competition.

Once candidate companions are confirmed, precise relative astrometry over time is mandatory to derive their orbital parameters (e.g., semi-major axis, eccentricity, and inclination), analyze dynamical properties for multi-planet systems (e.g., interactions, resonances, and stability), and in combination with radial velocity and/or absolute astrometric measurements, derive model-independent mass measurements. Current high-contrast extreme-adaptive optics imagers such as the Spectro-Polarimetric High-contrast Exoplanet REsearch (SPHERE)\cite{Beuzit2019}, the Gemini Planet Imager (GPI)\cite{Macintosh2014}, and the Subaru Coronagraphic Extreme Adaptive Optics and Coronagraphic High Angular Resolution Imaging Spectrograph (SCExAO+CHARIS)\cite{Jovanovic2015,Groff2016} explore the population of giant exoplanets and brown dwarf and stellar companions beyond typically 10 au, covering generally a small fraction of the orbit leading to degeneracies and biases in the orbital parameters ($<$20\% at the moment, because these instruments are available since only a few years). More precise measurements enable deriving more robust constraints on shorter timescales. The orbital elements can be compared to predictions from different formation scenarios for substellar companions (core accretion, gravitational instability in a circumstellar disk, and fragmentation of a protostellar disk) to constrain the formation mechanisms of the systems. Measuring the mass with model-independent methods is a fundamental step toward the calibration of models of the evolution of young giant planets, brown dwarfs, and low-mass stars (atmospheres and initial conditions).

Precise relative astrometry is also critical for predicting precise orbital positions for follow-up observations, e.g., at longer wavelengths\cite{Danielski2018} because of the lower angular resolution or with slit/fiber spectrometry\cite{Snellen2014,Lacour2019,Mawet2016,Otten2021}. Finally, precise relative astrometry is crucial for analyzing potential dynamical interactions in systems where a companion orbits in a circumstellar disk\cite{Delorme2017a, Maire2018} and for measuring slow motions of disk features (e.g., spirals, clumps, and shadows) to constrain the underlying production mechanism\cite{Boccaletti2015, Boccaletti2018, Ren2020}.

Precise and robust relative astrometric measurements over time in direct imaging require a good knowledge of the instrumental limitations and dedicated observing strategies. The typical astrometric precision of the first generation of exoplanet imaging instruments (e.g., the Nasmyth Adaptive Optics System Near-Infrared Imager and Spectrograph (NaCo) on the Very Large Telescope (VLT), the Near Infra Red Camera 2 (NIRC2) on the Keck telescope, the Near-Infrared Coronagraphic Imager (NICI) on the Gemini telescope) was of the order of 10~mas\cite{Chauvin2012, Konopacky2016a, Biller2013}. Dedicated procedures were developed for the first dedicated exoplanet imaging instruments SPHERE and GPI to minimize the systematic astrometric error budgets to reach precisions down to $\sim$1--2~mas not counting the noise contribution\cite{Macintosh2014, Beuzit2019}. Such precise measurements are more sensitive to previously neglected systematic uncertainties due to biases in the data analysis and/or calibration and our limited knowledge of the thermo-mechanical stability of the instruments. For instance, the astrometric calibration of GPI was recently revised after the correction of issues in the data reduction pipeline and in the data calibration\cite{DeRosa2020}. A homogeneous astrometric strategy is mandatory for precise relative astrometry over time because it minimizes potential systematic uncertainties in the calibration and enables the analysis of its stability over time. A stable astrometric calibration reduces the overhead at the telescope by relaxing the need to take nighttime calibration data close in time to the science observations. Maximizing the observing efficiency and optimizing the data exploitation are important in a context of high pressure on the current 8--10~m class telescopes and the even higher pressure expected for the upcoming extremely large telescopes.

The SPHERE consortium was granted with a guaranteed-time program of 230 nights over five years, from which 200 nights are dedicated to the SpHere INfrared survey for Exoplanets (SHINE) program to detect and characterize in the near-infrared (near-IR) the population of young giant planets and brown dwarfs at wide orbits ($\gtrsim$5~au)\cite{Desidera2021,Langlois2021,Vigan2021}. One of the major goals of the survey was to measure precisely the orbit of the detected companions. A dedicated and homogeneous strategy for astrometry was designed. A first analysis of the astrometric calibration of SPHERE was performed using data taken in the first two years of operations\cite{Maire2016b} and showed promising results for the calibration stability.

We present in this paper an analysis of SPHERE astrometric data obtained over five years. We derive updated estimates for the astrometric calibration parameters and confirm their stability. We first recall the astrometric strategy defined for SPHERE (Sec.~\ref{sec:strat}). We then describe issues encountered during the course of the survey analysis (Sec.~\ref{sec:issues}). We also present an updated analysis of the SPHERE astrometric data (Sec.~\ref{sec:astrocal}). Finally, we outline key lessons learned for high-precision relative astrometry with SPHERE for optimizing the preparation of the exploitation of the next generation of exoplanet imaging instruments especially for the Extremely Large Telescope (ELT) being built by the European Southern Observatory (ESO, Sec.~\ref{sec:lessons}).

\section{Astrometric strategy for the VLT/SPHERE exoplanet imager}
\label{sec:strat}

\begin{figure}[t]
\centering
\includegraphics[width=.85\textwidth]{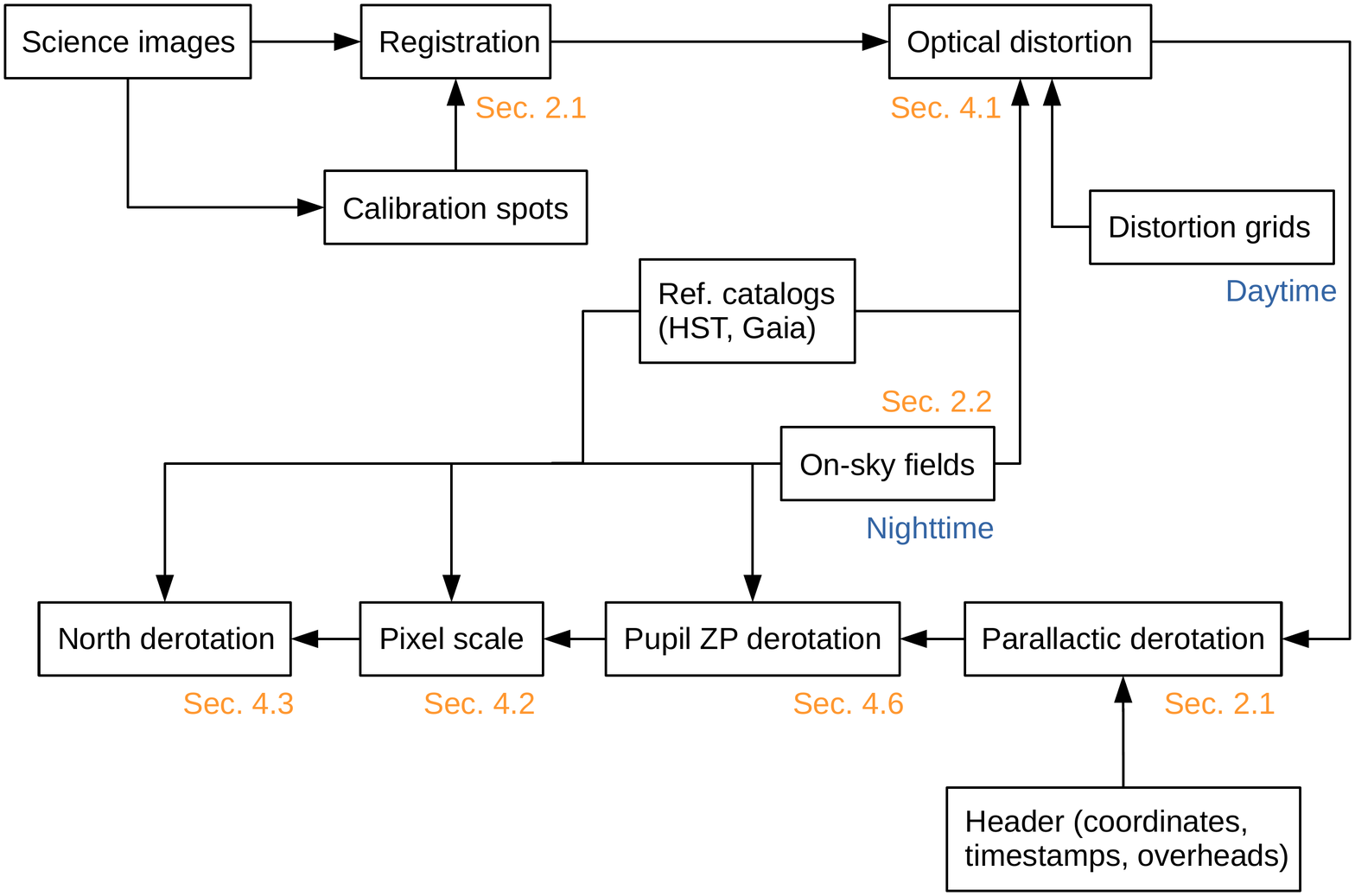}
\caption{\rev{Block diagram of the astrometric calibration of SPHERE/IRDIS data.}}
\label{fig:blockdiagrastrocal}
\end{figure}

\rev{Figure~\ref{fig:blockdiagrastrocal} shows the block diagram of the astrometric calibration strategy for SPHERE/IRDIS data. The steps are discussed in this section and Sec.~\ref{sec:astrocal}.}

\subsection{Exoplanet imaging with VLT/SPHERE and astrometric requirements}
\label{sec:sciobs}

Exoplanet imaging with SPHERE mainly uses its extreme adaptive optics (AO) system\cite{Fusco2014} with the infrared dual-band imager and spectrograph (IRDIS)\cite{Dohlen2008a} and the near-IR integral field spectrograph (IFS)\cite{Claudi2008}. IRDIS (FoV 11$''$\,$\times$12.5$''$) offers dual imaging with broad-band and narrow-band filters covering different bandpasses over the $Y$ to $Ks$ bands (0.95--2.32~$\mu$m)\cite{Vigan2010}. IFS (FoV 1.73$''$\,$\times$1.73$''$) can simultaneously observe the $YJ$ bands (0.95--1.35~$\mu$m, $R$\,$\sim$\,54) or the $YJH$ bands (0.95--1.65~$\mu$m, $R$\,$\sim$\,33). IRDIS can be operated alone, but IFS can only be operated in parallel with IRDIS.

To attenuate the stellar light contamination, both instruments use coronagraphy\cite{Boccaletti2008c}. The focal plane masks are located in the common optical relay of SPHERE and are thus common to both instruments. They are optimized over their wavelength range. The Lyot stops are located inside the instruments themselves. Both instruments are also operated in pupil-tracking mode to take advantage of the angular differential imaging (ADI) technique\cite{Marois2006a} to further suppress the stellar contamination. As a result, the FoV around the targeted star, including potential point sources, rotates. After the removal of the stellar contamination, the individual images are realigned and combined to detect faint point sources close to and with limited pollution from the star.

The SPHERE requirements on the measurements of the separation and position angle of detected point sources are 5~mas and 0.2$^{\circ}$, respectively (goal 1~mas and 0.1$^{\circ}$). Extensive tests using injections of synthetic point sources in laboratory data processed with spectral differential imaging\cite{Racine1999, Sparks2002} showed that for separation measurements, astrometric accuracy is better than 1.5--2~mas over a FoV of 1.6$''$ for detections at signal-to-noise ratios (S/N) above 10\cite{Zurlo2014}, within the requirements.

\begin{figure}[t]
\centering
\includegraphics[width=.84\textwidth]{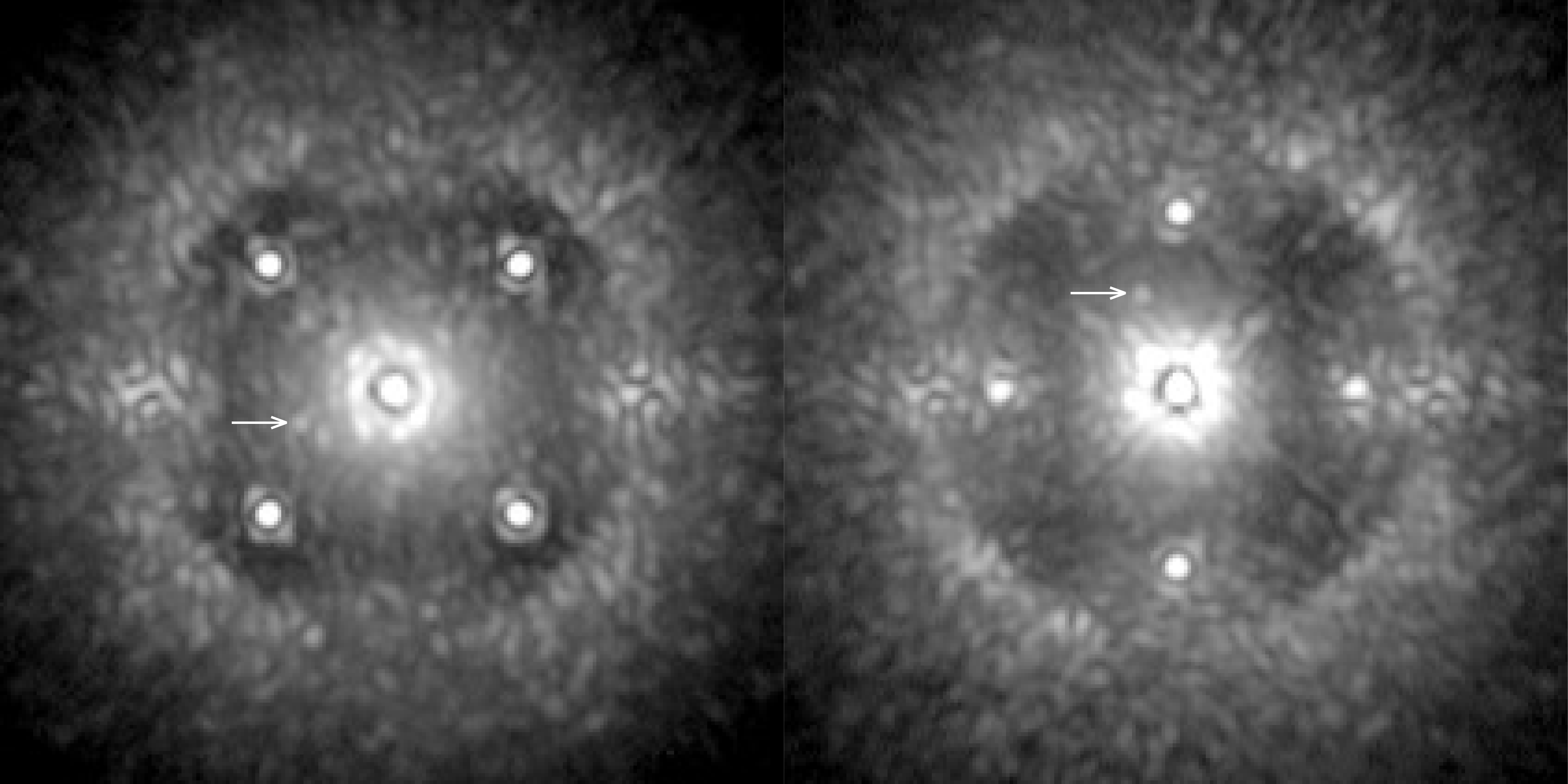}
\caption{Example of science images obtained with SPHERE/IRDIS after an integration time of 4\,s, taken on $\beta$ Pictoris on 2015 February 5 UT\cite{Lagrange2019}. The four calibration spots used to determine the location of the star behind the coronagraph can be seen on a square pattern (left) and a cross pattern (right) centered on the star. The AO correction radius is visible as a bright ring further out. The white arrow indicates the giant planet $\beta$ Pictoris b\cite{Lagrange2010b} (separation $\sim$0.33$''$). The same filter+coronagraph setup was used but the left image was taken in better atmospheric conditions, hence the darker appearance of the area within the AO correction radius (``AO dark hole'').}
\label{fig:exampleim}
\end{figure}

An efficient attenuation of the stellar residuals and high-precision relative astrometry critically depend on a precise estimate of the location of the star behind the coronagraph\cite{Marois2006b, Sivaramakrishnan2006} and on its stability. For SPHERE, a calibration image is recorded for this purpose at the beginning and the end of an observing sequence with four crosswise calibration spots (Fig.~\ref{fig:exampleim}). The calibration spots are produced by applying a periodic modulation on the AO deformable mirror\cite{Langlois2013}. To maximize the S/N of the detected point sources and measure precise position angles, a precise azimuthal realignment of the individual images before their combination is mandatory.

To correct the SPHERE coronagraphic images for cosmetic defects, dithering the star on the detector during the observations is not possible because the \rev{position of the coronagraph in the optical path} is fixed. This is useful for bad pixel correction but was foreseen first to improve the flat accuracy to 0.1\%. Instead, the IRDIS detector is dithered thanks to a dithering stage on which it is mounted. However, the dithering stage has a finite positioning accuracy of 0.74~mas\cite{Vigan2016}, which has to be taken into account in the astrometric error budget when the calibration spots are not used during the whole science observation.

\subsection{Astrometric observations}
\label{sec:astrobs}

The astrometric calibration strategy for the SPHERE SHINE survey was conceived before the instrument commissioning in 2014 and subsequently refined. Regular observations of astrometric fields were performed to derive the optical distortion, pixel scale, and orientation to the North of the images. A regular monitoring is mandatory to measure potential small variations in the parameters, due to opto-mechanical variations (e.g., positioning accuracy of the image derotator) or technical interventions on the instrument using in complement the distortion grid. The homogeneous strategy has significantly facilitated astrometric studies with SPHERE both in the SHINE survey and in open-time programs.

\subsubsection{\rev{Selection criteria for the astrometric fields and catalogs}}
\rev{Several criteria have to be considered when selecting astrometric fields. Ground-based observations require at least two fields to cover the whole year of observations because a given field is best observable for about 6 months.}

\rev{Deriving the optical distortion needs fields with a high density of stars but that can be well separated with a homogeneous distribution in the FoV.} Stellar binaries and multiples have been commonly used to calibrate high-contrast imaging surveys\cite{Hayward2014, Maire2015, Chauvin2015}. However, they only allow for measuring the pixel scale and orientation to the North of the images. Distortion grids in the instruments have been used for measuring the optical distortion of the images, but they can measure the distortion due to the instrument optics only and do not include the telescope. For the SPHERE SHINE survey, we chose to observe fields in stellar clusters because the large number of stars available allows for more precise measurements. They also allow for measuring the distortion from the telescope optics. Nevertheless, in the case of the VLT, the optical distortion is expected to be small with on-sky measurements of the Galactic Center with NaCo indicating distortion effects below 0.1~mas over a 5.4$''$ FoV\cite{Trippe2008}.

Minimizing the integration times needed for the observations requires fields with bright stars but which will not saturate the detector (for SPHERE, stars with $H$\,$<$\,8.6~mag will saturate the detectors at the shortest integration times, 0.83s for IRDIS and 1.66s for IFS). Neutral density filters can be used to avoid saturation but they will slightly affect the measured distortion pattern compared to the science observations for which they are not used. \rev{Nevertheless, the neutral density filters are not conjugated to a focal plane so we do not expect a significant effect on the measured distortion. We} did not use neutral density filters for the astrometric observations in the SPHERE SHINE survey\footnote{Most observations were obtained in narrow-band filters and with a coronagraph, so neutral density filters were not needed. \rev{The ESO calibration data (Sec.~\ref{sec:esocalib}) taken with the neutral density filters could be used to assess their effect on the measured distortion, but unfortunately, most of the} \rev{ESO data obtained on stellar clusters so far suffer from a sensitivity issue. The ESO calibration plan has been improved, so such an analysis would be feasible in the future.}}.

\rev{The use of AO imposes additional constraints on the astrometric fields} with the presence of a bright star for guiding ($R\lesssim$13.5 mag for SPHERE). This strongly limits the selection of fields within stellar clusters. Bright stars are not used for the calibration because they are saturated in the data typically used for catalog positions (e.g., Hubble Space Telescope, hereafter HST).

\rev{Finally, the} choice of the reference catalog for the stellar positions is important. It should be precise, preferably provide the orbital or proper motions of the individual stars, have a good absolute calibration, and obtained with an instrument with a similar angular resolution to the observations to be calibrated. 

\subsubsection{\rev{Astrometric fields used}}
\label{sec:astrometricfields}
\begin{figure}[t]
\centering
\includegraphics[width=.84\textwidth]{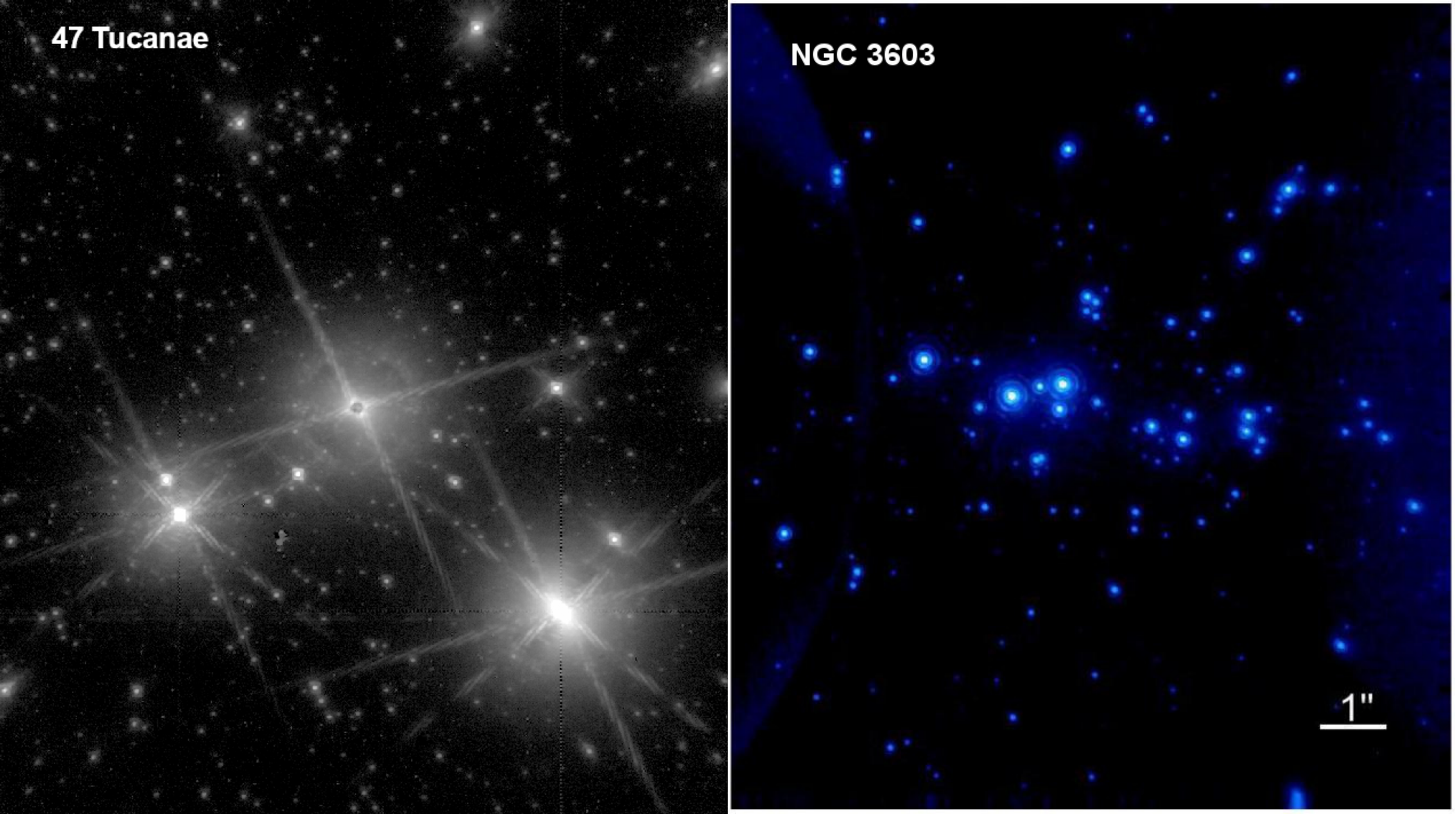}
\caption{SPHERE/IRDIS images of the main astrometric calibration fields, 47 Tucanae (left) and NGC3603 (right). North is up and East toward the left.}
\label{fig:astrocalfields}
\end{figure}

For calibrating the SPHERE SHINE survey, we selected fields in the globular stellar cluster 47~Tuc and the open stellar cluster NGC3603 as main calibrators (Fig.~\ref{fig:astrocalfields}). Both catalogs are based on HST data, which are precise and have a good absolute calibration. To mitigate (partially) the poorer angular resolution of the HST compared to the VLT due to the smaller telescope aperture, HST data obtained in the visible were used. 

The catalog for 47~Tuc contains the proper motions (precision 0.3~mas/yr, reference epoch 2006.20; priv. comm. from A. Bellini/STScI, see Ref.~\citenum{Bellini2014} for the methodology), but not the catalog for NGC3603\cite{Khorrami2016}. For this reason, we selected 47~Tuc as the reference calibration field. However, NGC3603 is located at a larger distance from the Sun than 47~Tuc ($\sim$9.5~kpc\cite{CantatGaudin2018} vs. 4.45~kpc\cite{SChen2018}) and has a low mass and concentration with respect to typical globular clusters ($\sim$10$^4$~$M_\odot$ vs. $\sim$10$^6$~$M_\odot$ for 47~Tuc\cite{Harayama2008,HenaultBrunet2020}), hence the internal velocity dispersion is expected to be smaller. Therefore, stellar proper motions should have smaller effects on the precision of the catalog positions. In the future, we envision to use the multi-epoch SPHERE data set obtained on NGC3603 to derive the internal proper motions within this cluster (using the first SPHERE epoch obtained as a reference) and include this information in the catalog\cite{Khorrami2021}.

Other calibration fields which were occasionally observed include the $\theta^1$ Ori  Trapezium B1–B4\cite{Close2012} and NGC6380 (reference catalog obtained through priv. comm. from E. Noyola/Univ. Texas Austin). \rev{For the astrometric calibration derived using the $\theta^1$ Ori  Trapezium B1--B4, we accounted for systematic uncertainties of 0.030~mas/pix for the pixel scale and 0.062$^{\circ}$ for the offset angle to the North\cite{Close2012}. The stars B2, B3, and B4 were not used for the calibration. We used instead the star B1 and two stars located far in the IRDIS FoV (separation $>$6$''$, see Fig.~1 in Ref.~\citenum{Maire2016b}). We neglected stellar differential motions since the observations of Ref.~\citenum{Close2012}.}

\subsubsection{\rev{Observing setup}}
The astrometric fields are mainly observed with a coronagraph in field-tracking mode, where the FoV is stabilized. This makes combining images easier to get more precise measurements, when the calibration stars are faint and/or for measuring the optical distortion. Measuring the optical distortion in pupil-tracking mode is not as straightforward as in field-tracking mode. The image derotator is the first component in the optical train of the instrument, hence the distortion pattern does not rotate in pupil-tracking mode. Thus, pupil-tracking images have to be corrected for the distortion before being realigned and combined. Measuring the optical distortion in such data would require astrometric fields with a large number of stars detected with high signal-to-noise ratios in single short-exposure images. To measure the angle offset between the pupil-tracking and field-tracking modes due to the zeropoint angle of the SPHERE pupil (Sec.~\ref{sec:ptftoffset}), we observed a same astrometric field in both modes consecutively at several epochs.

When observing the $\theta^1$ Ori  Trapezium B1–B4 and NGC3603, the AO guide star is offset out of the coronagraphic mask using a tip-tilt mirror so that it can be used in the calibration (because it is not saturated).

\subsubsection{\rev{Verification of the absolute astrometric calibration}}
To verify the absolute astrometric calibration of SPHERE, Ref.~\citenum{Langlois2021} compared the relative astrometry for 7 widely-separated ($\gtrsim$5$''$) and bright candidate companions observed with SPHERE and present in the Gaia DR2 catalog. A direct comparison to 47~Tuc and NGC3603 is not possible because these fields are typically crowded in Gaia data due to the poorer angular resolution. The Gaia DR2 position offset compared to SPHERE averaged over the sample is -2.8$\pm$1.5~mas (3.9~mas rms\footnote{\rev{The uncertainty on the mean value is the rms error divided by the square root of the number of measurements.}}) in separation and 0.06$\pm$0.04~deg (0.11~deg rms) in position angle. The rms agree well with the expected uncertainties in these quantities in SPHERE data. Ref.~\citenum{Bonavita2021} performed a similar analysis using the Gaia EDR3 catalog. For 12 physical binary systems and accounting for the relative proper motion between the two components between the Gaia EDR3 and SPHERE epochs, they find an average difference in the separation and position angle between Gaia EDR3 and SPHERE measurements of $1.6\pm0.8$~mas (rms=2.8~mas) and $-0.12\pm0.03$~deg (rms=0.11~deg), respectively. The small zero point offsets in scale and position angle are well within the uncertainties of the SPHERE astrometric calibration\cite{Maire2016b}. The comparison with Gaia indicates that the accuracy of SPHERE astrometry is better than 3 mas even at large separation.

\section{Issues encountered}
\label{sec:issues}

We review in this section technical issues that we encountered during the course of the SPHERE operations and how we handled them.

\subsection{SPHERE time reference}
\label{sec:timerefissue}
From December 2015 to 2016 February 7 UT, SPHERE suffered from variations up to $\sim$1$^{\circ}$ in the offset angle to the North direction within timescales of a few days\cite{Maire2016b}. This issue also affected the measurement of the position angles of the physical sources detected in the science images. These variations were caused by errors in the reference position of the image derotator. Dome-tracking tests with the internal distortion grid of SPHERE showed that they were caused by synchronization issues between the SPHERE and VLT internal clocks. These issues were in fact present since the first light of SPHERE but somewhat mitigated by more frequent instrument resets. The derotation error of the image derotator could be predicted using the information in the data headers and the comparison with the astrometric observations confirmed the predictions.

The issue affects both field-tracking and pupil-tracking data. The formula provided in Ref.~\citenum{Maire2016b} is valid for field-tracking data. For pupil-tracking data, the formula is:
\begin{equation}
\epsilon\,=\,atan \left( tan \left( (ALT_{START}-2 \times DROT2_{BEGIN}) \times \frac{\pi}{180} \right) \right) \times \frac{180}{\pi}
\end{equation}
\rev{where $\epsilon$ is the derotation error, $ALT_{START}$} the telescope altitude at the beginning of the observations provided by the telescope control software, and $DROT2_{BEGIN}$ is the position angle of the SPHERE derotator at the beginning of the observations calculated by the SPHERE lighting control unit. \rev{To correct for the derotation error, $\epsilon$ should be added to the uncorrected North correction angle (Sec.~\ref{sec:northangle} and Fig.~4 in Ref.~\citenum{Maire2016b}).} 

The ability to predict and correct for this issue was important to optimize the use of the data taken during the first two years of operations for high-precision astrometry. A component was installed in the SPHERE lighting control unit on 2016 July 13 UT for ensuring its proper synchronization with the telescope internal clock every day. Subsequent measurements of the correction angle to the North confirm that the clock drift issue is solved (Sec.~\ref{sec:northangle}). The correction of this issue emphasizes the importance of having detailed information recorded in the data headers.

\subsection{Backlash of the image derotator}
\label{sec:derotbacklash}
The image derotator has been shown to suffer from backlash\cite{Beuzit2019}. Jumps of $\sim$0.05$^{\circ}$ are measured near the meridian passage. This issue affects the measurement of the position angle of the sources detected in the science images. Investigations are ongoing to determine if the backlash could be predicted and corrected. This issue likely mainly accounts for the uncertainty derived for the angle offset between the pupil-tracking and field-tracking observing modes (Sec.~\ref{sec:ptftoffset}).

\subsection{Instability of the star centering during science sequences}
\label{sec:starceninstability}
\begin{figure}[t]
\centering
\includegraphics[width=.8\textwidth, trim = 0mm 0mm 0mm 10mm,clip]{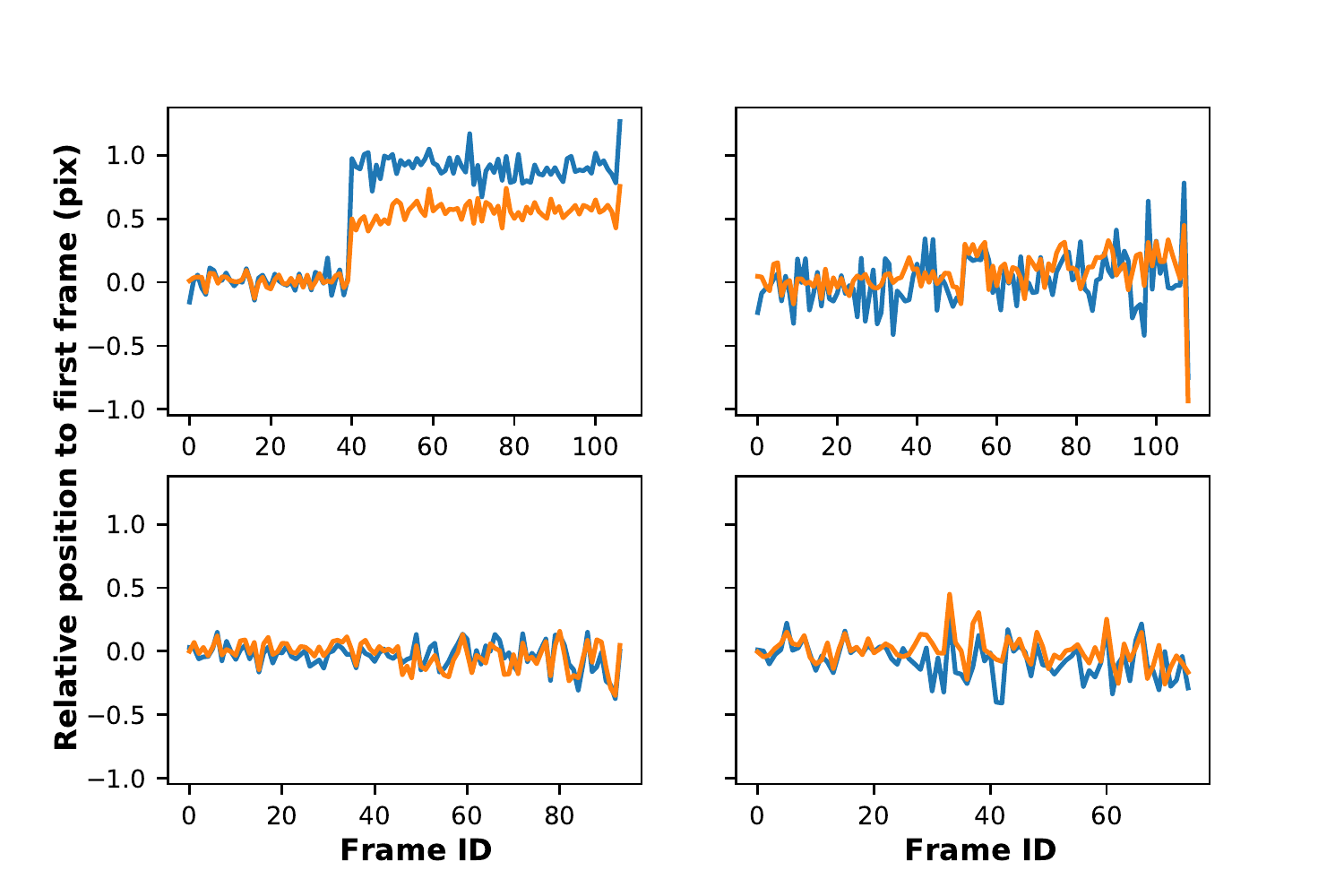}
\caption{Evolution as a function of the frame ID of the star location relative to the position measured in the first frame in the sequence measured in a few science sequences where the calibration spots were used (in pix, 0.1~pix=1.25~mas). Blue: $x$ direction, orange: $y$ direction.}
\label{fig:starcentinstab}
\end{figure}

We noted instabilities in the star centering during \rev{some science sequences by measuring the frame by frame evolution of the image center using the calibration spots. These centering instabilities} can amount up to $\sim$2--3~mas. \rev{The SHINE science observations were mostly performed without the calibration spots.} If \rev{centering instabilities are present in a science sequence where the calibration spots were not used simultaneously, they cannot be measured and corrected and will affect} the quality of the frame registration, the detection performances of ADI algorithms, and the astrometric accuracy on the measurements of the detected sources. Figure~\ref{fig:starcentinstab} shows a few examples of instabilities measured in science sequences where the calibration spots to monitor the star location were used simultaneously. Different variations can be seen: random variations, drifts, jumps.

Random variations can be attributed to the precision at which the SPHERE differential tip-tilt sensor\cite{Baudoz2010} can maintain the star on a given position. Drifts and jumps are more difficult to explain. \rev{The jitter (jumps) effect comes from the AO residual tip-tilt correction while the drifts could be most likely attributed to the differential tip-tilt sensing control loop. Because the atmospheric dispersion corrector is not located in a plane conjugated to the focal plane its effect on the distortion is negligible\cite{Pathak2016}. It could create optical aberrations but most of them will be corrected by the AO system.}

To mitigate the error term due to the frame registration in the astrometric error budget, we modified our strategy for orbital monitoring observations of confirmed companions to use the calibration spots simultaneously. Using the calibration spots also minimizes the error term due to the detector dithering (Sec.~\ref{sec:sciobs}).

\section{5-yr analysis of SPHERE/IRDIS astrometric calibration data}
\label{sec:astrocal}

\begin{table}[t]
\centering
\caption{Observing setup for observations of astrometric fields with coronagraph. \rev{DIT is the Detector Integration Time and Nfr the number of frames. For the Nfr column, the two numbers are the number of frames in a single data cube and the number of data cubes recorded in the sequence, respectively.}}
\label{tab:astrocalsetups}
\begin{tabular}{cccccc}
\hline
Field & Filter & DIT & Nfr \\
 & & (s) & \\
\hline
47Tuc & DB\_H23 & 8 & 4$\times$16 \\
47Tuc & DB\_K12 & 4 & 5$\times$16 \\
47Tuc & BB\_H & 5 & 2$\times$8 \\
NGC3603 & DB\_H23 & 8 & 30$\times$3 \\
NGC3603 & DB\_K12 & 4 & 40$\times$3 \\
NGC3603 & DB\_J23 & 4 & 30$\times$3 \\
NGC3603 & BB\_J & 2 & 80$\times$5 \\
NGC3603 & BB\_Ks & 0.83 & 80$\times$5 \\
$\theta^1$Ori B1-B4 & DB\_H23 & 8 & 16$\times$2 \\
$\theta^1$Ori B1-B4 & DB\_K12 & 4 & 60$\times$3 \\
NGC6380 & DB\_H23 & 16 & 4$\times$16 \\ 
NGC6380 & DB\_K12 & 16 & 4$\times$16 \\
\end{tabular}
\end{table}

Astrometric observations have been obtained regularly since SPHERE has been in operation. SPHERE was commissioned between May and October 2014 and science verification observations took place between December 2014 and March 2015. SPHERE/SHINE observations started in February 2015. SPHERE has been offered to the community since April 2015. Table~\ref{tab:astrocalsetups} provides the setup used for the observation of a given field in a given filter.

To analyze the astrometric data and derive the calibration, we developed a tool, which requires minimum interactions from the user\cite{Maire2016b}. It has been subsequently included in the SPHERE Data Center\cite{Delorme2017b}\footnote{https://sphere.osug.fr/spip.php?rubrique16} and used to analyze homogeneously all the astrometric data obtained in the SHINE survey and by the ESO Staff as part of the monthly ESO calibration plan \rev{(Sec.~\ref{sec:esocalib})}. A calibration table has been produced for the science observations obtained in the SHINE survey and another table for those obtained by open-time programs. The calibration tables are used by the SPHERE Data Center pipeline to calibrate the data according to the setup used and closeliness in time.

Briefly, the astrometric data were reduced with the SPHERE data reduction pipeline\cite{Pavlov2008} either with a custom script or in the SPHERE Data Center. They were subsequently analyzed with custom IDL routines to derive the calibration. For data obtained in field-stabilized mode, the individual frames are first selected based on the flux statistics and combined to enhance the signal-to-noise ratio of the detected stars. Then, the positions of the stars are measured with Gaussian fitting using the mpfit library\cite{Markwardt2009}. \rev{We compared the results obtained with Moffat fitting and centroiding through derivative search (using the cntrd routine of the astron library\cite{Landsman1993}) and did not find significant systematics.} The counter-identification between the SPHERE positions and the catalog positions is done using estimates for the separations and the position angles assuming approximate values for the pixel scale and North offset correction angle (plus the IFS angle offset relative to IRDIS for IFS observations) and tolerance criteria. After the counter-identification, the average pixel scale and North offset correction angle are derived from the statistics of all the available stellar pairs after removing outliers using sigma clipping. The uncertainties on the values are the standard deviations of the values measured for the stellar pairs. They were chosen to be conservative to include potential variations due to changes within the instrument over the duration of the SHINE survey runs, which were typically of a few days. For $\theta^1$ Ori B1--B4, the uncertainties include in addition systematic uncertainties \rev{(Sec.~\ref{sec:astrometricfields})}. For correcting the optical distortion, two options are available, either measuring the actual distortion (relevant if several tens of stars are detected in the FoV) or applying a generic correction determined from on-sky data taken during the instrument commissioning. The on-sky optical distortion is measured by fitting linear coordinate transformations between the catalog and the SPHERE positions. All angles provided in the following sections are counted positive from North to East. For the IRDIS data, we force the use of the same stars in the left and right dual fields on the detector.

\begin{figure}[t]
\centering
\includegraphics[width=.92\textwidth]{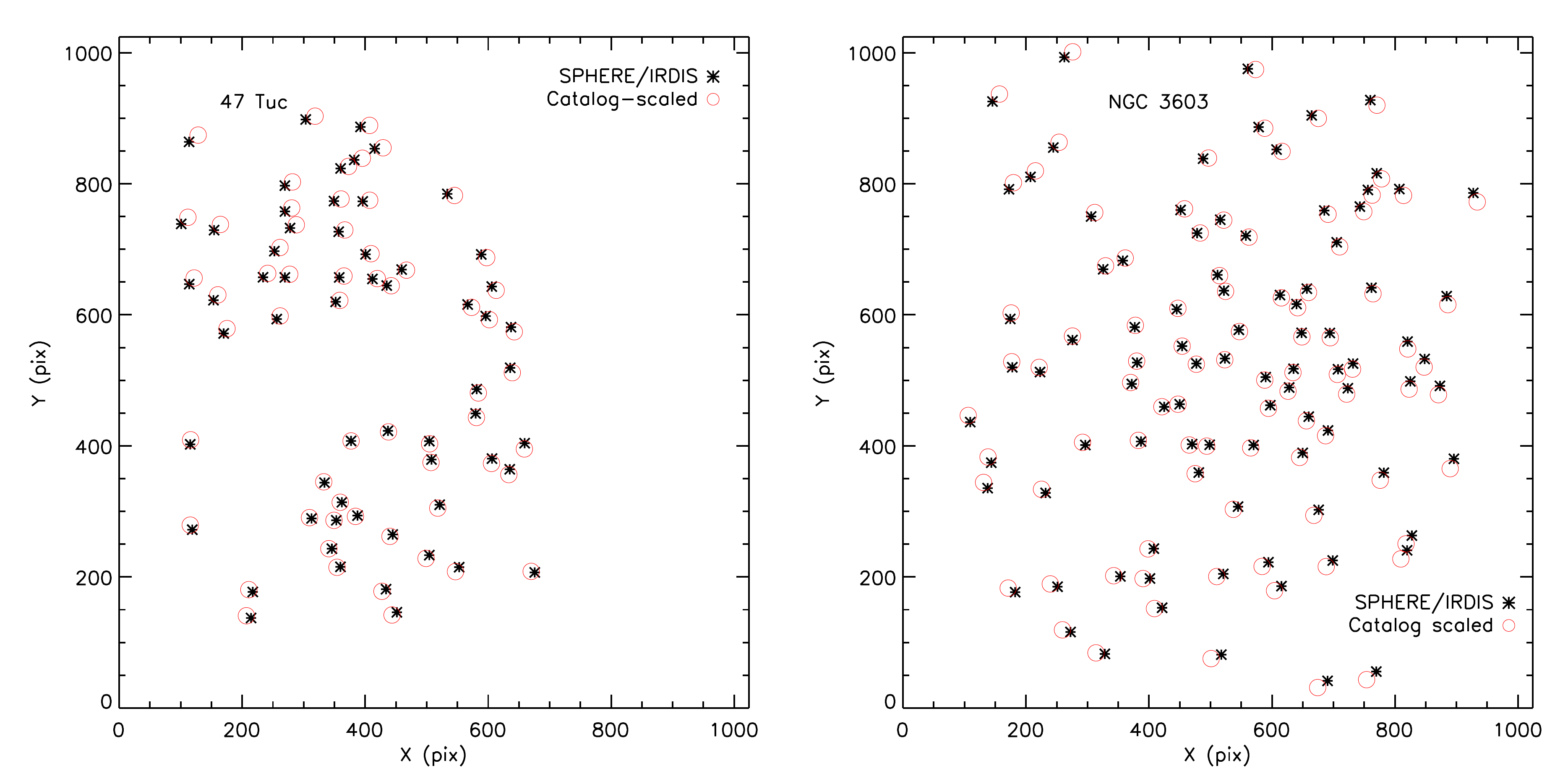}
\caption{Comparison of the measured positions corrected for the distortion and of the catalog positions scaled and translated for 47 Tuc (left) and NGC 3603 (right). \rev{The measured stellar patterns are rotated compared to the catalog stellar patterns because they were not corrected for the rotation to the North. To align SPHERE images with North up, they should be rotated in the clockwise direction (i.e., the correction angle to the North is negative, see Sec.~\ref{sec:northangle}).}}
\label{fig:astrocalpos}
\end{figure}

Figure~\ref{fig:astrocalpos} compares the measured positions corrected for the optical distortion (Sec.~\ref{sec:anamorphism}) and the catalog positions scaled and translated for the 47~Tuc and NGC3603 fields. The measured stellar pattern is rotated counterclockwise with respect to the catalog stellar pattern \rev{because we did not correct the measured positions for the rotation to the North. The images should be rotated clockwise to align them with North up (i.e., the correction angle to the North is negative, Sec.~\ref{sec:northangle})}. The empty regions in the diagram for 47~Tuc are partly due to the very bright stars present in the FoV (Fig.~\ref{fig:astrocalfields}) and which are saturated in the HST data.

\rev{Assuming that a SPHERE/IRDIS image is corrected for the optical distortion and for the parallactic rotation for pupil-tracking observations or for the position angle of the image derotator for field-tracking observations, the on-sky position angle is related to the position angle measured in the image by:}
\begin{equation}
 PA_\mathrm{SKY} = PA_\mathrm{IRD} + \mathrm{corr.~angle~pupil~zeropoint} + \mathrm{corr.~angle~to~North}
\end{equation}
\rev{for pupil-tracking observations, and}
\begin{equation}
PA_\mathrm{SKY} = PA_\mathrm{IRD} + \mathrm{corr.~angle~to~North}
\end{equation}
\rev{for field-tracking observations,}\\
\rev{with $PA_\mathrm{SKY}$ the on-sky position angle, $PA_\mathrm{IRD}$ the position angle measured in the SPHERE/IRDIS image, corr. angle pupil zeropoint = 136.00$^{\circ}$ (Sec.~\ref{sec:ptftoffset}), and corr. angle to North = -1.76$^{\circ}$ (Sec.~\ref{sec:northangle}).}

\begin{figure}[t]
\centering
\includegraphics[width=.49\textwidth]{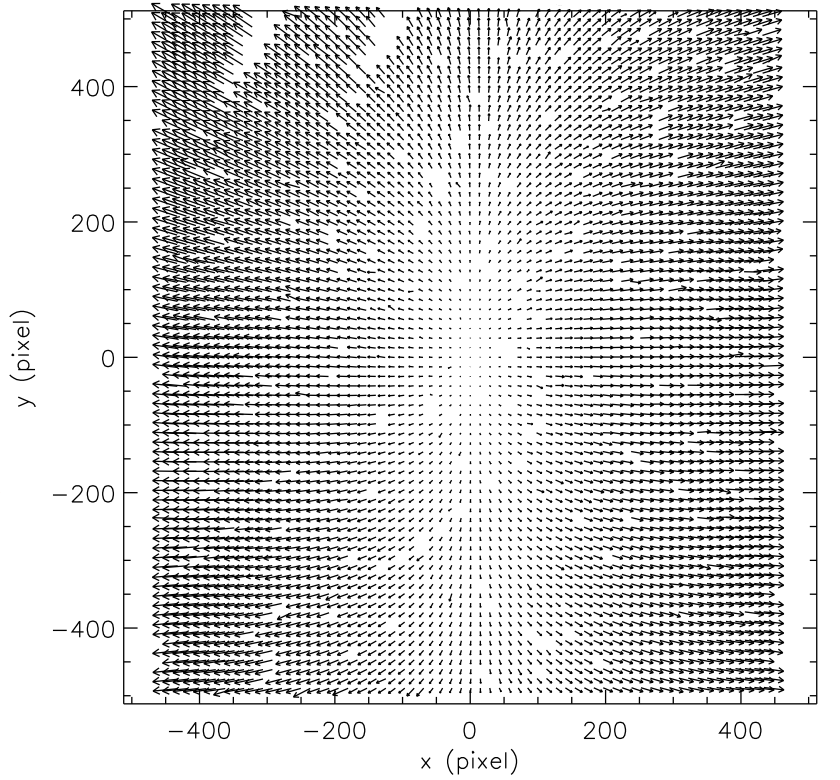}
\caption{\rev{SPHERE/IRDIS distortion map obtained using the internal distortion grid. The distortion vectors are magnified by a factor 10 for readibility.}}
\label{fig:distmapgrid}
\end{figure}

\subsection{\rev{Optical anamorphism}}
\label{sec:anamorphism}

The data analysis of the 47~Tuc field observed during the SPHERE commissioning showed that the optical distortion of the images is dominated by the distortion of the optics in SPHERE\cite{Maire2016b}. Laboratory measurements confirmed that the cylindrical mirrors in the SPHERE common path are the main source for the instrument optical distortion. The distortion manifests in the differences in the pixel scale between the horizontal and vertical directions of the IRDIS detector of 0.60$\pm$0.02\%, i.e. 6~mas at 1$''$. This is larger than the SPHERE astrometric requirements (5~mas). To correct the SPHERE data for the optical distortion, we have relied on the on-sky measurements. We did not use the distortion grid data because of issues in the first months of the survey with the data quality (saturation, see below) and with the data reduction recipe and also because the distortion patterns measured on sky and using the internal calibration data are similar. The SPHERE data reduced at the SPHERE Data Center are corrected for the pixel scale difference (the raw IRDIS images are vertically stretched by a factor 1.006 on both image sides). Using 47~Tuc data, we measured that the average residual error over the full IRDIS FoV of this first-order correction compared to higher-order corrections is below 1~mas, i.e. 0.09~pixel. The distortion pattern is not affected by the tracking mode of SPHERE (tracking of the instrument pupil or of the on-sky field) since the instrument image derotator is the first element in the optical train. The distortion is common to all SPHERE science subsystems except for a rotation of their respective FoVs and is stable over time.

Using a more extended set of on-sky data taken during the SHINE survey, we reassessed the ratio of the horizontal and vertical pixel scales for the dual-band DB\_H23 and DB\_K12 filter configurations with coronagraph. For this analysis, we used NGC3603 and not 47~Tuc because of the more homogeneous distribution of the stars in the FoV (Fig.~\ref{fig:astrocalpos}). Based on six NGC3603 data sets taken in 2017-2018 in the DB\_H23+N\_ALC\_YJH\_S configuration, we find an average pixel scale ratio of 1.0070$\pm$0.0005, which is larger by $\sim$1.9$\sigma$ than the value derived using the commissioning data. The analysis of five NGC3603 data sets taken in 2017-2018 in the DB\_K12 filter gives an average pixel scale ratio of 1.0072$\pm$0.0003.

\begin{figure}[t]
\centering
\includegraphics[width=.9\textwidth]{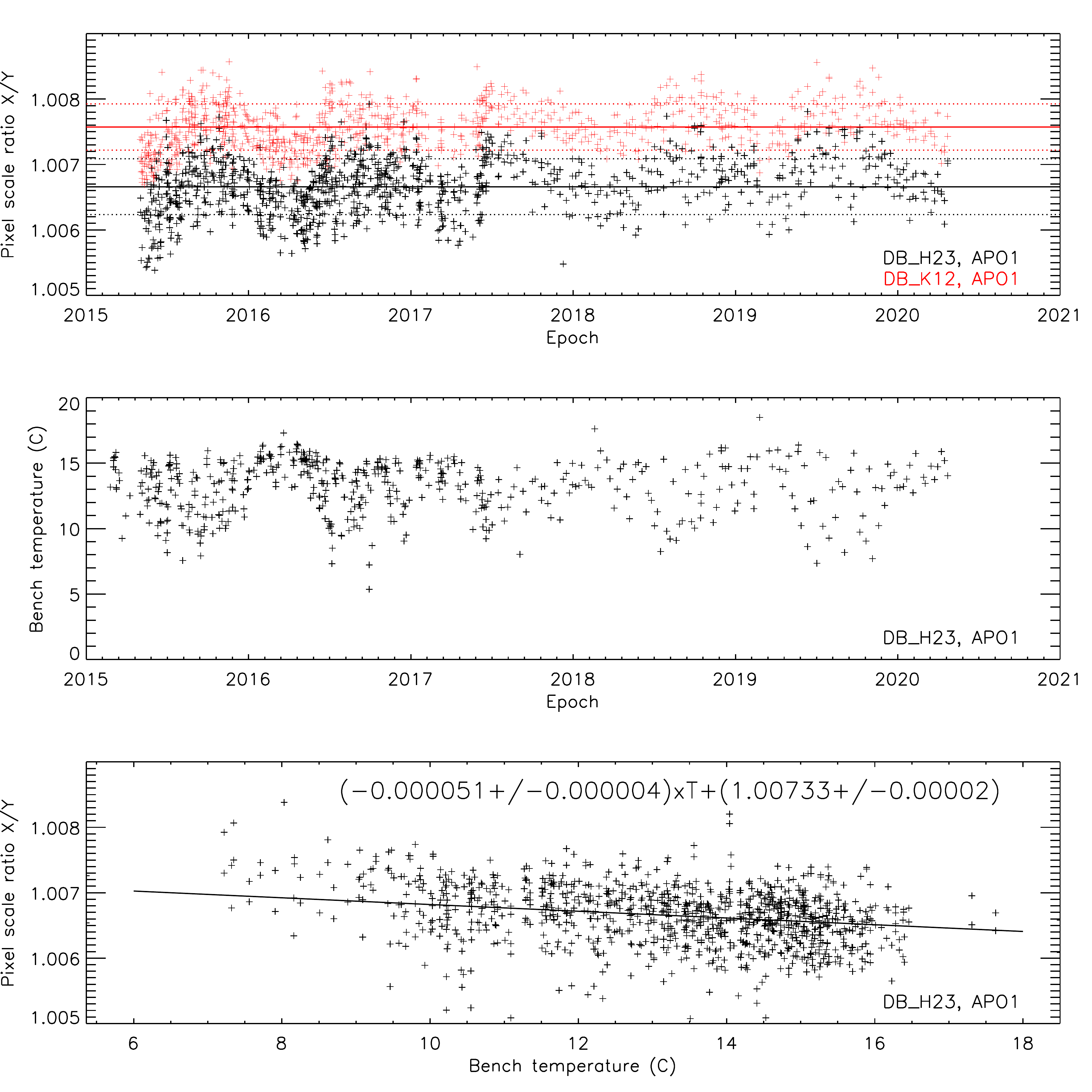}
\caption{Temporal evolution of the pixel scale ratio between the horizontal and vertical directions of the SPHERE/IRDIS detector for the dual-band configurations with coronagraph for the DB\_H23 filter and for the DB\_K12 filter (top), the evolution of the bench temperature as a function of time (middle), and evolution of the pixel scale ratio between the horizontal and vertical directions of the SPHERE/IRDIS detector as a function of the bench temperature (bottom). In the top panel, the solid lines indicate the mean values and the dotted lines the 1$\sigma$ intervals. In the bottom panel, the solid line shows a linear fit to the data.}
\label{fig:sphanamorphtime}
\end{figure}

\begin{table}[t]
\centering
\caption{Mean and standard deviations measured for the pixel scale ratio between the horizontal and vertical directions of the SPHERE/IRDIS detector as a function of the coronagraph+filter configuration.}
\label{tab:anamorphfilt}
\begin{tabular}{ccccc}
\hline
Filter & DB\_H23 & DB\_H23 & DB\_K12 & DB\_K12 \\
Coronagraph apodizer & APO1 & CLEAR & APO1 & CLEAR \\
\hline
Pixel scale ratio & 1.0067 & 1.0066 & 1.0076 & 1.0075 \\
& $\pm$0.0004 & $\pm$0.0005 & $\pm$0.0004 & $\pm$0.0004 \\
\hline
Filter & BB\_Y & BB\_J & BB\_H & BB\_Ks \\
Coronagraph apodizer & CLEAR & CLEAR & CLEAR & CLEAR \\
\hline
Pixel scale ratio & 1.0076 & 1.0075 & 1.0074 & 1.0075 \\
& $\pm$0.0004 & $\pm$0.0003 & $\pm$0.0004 & $\pm$0.0003 \\
\hline
\end{tabular}
\end{table}

As a cross-check of the on-sky results, we used the systematic reductions of the internal distortion grid data performed by the Quality Control Group at ESO Garching\footnote{\url{https://www.eso.org/observing/dfo/quality/SPHERE/reports/FULL/trend_report_IRDIS_DBI_DIST_H_FULL.html} for the dual-band imaging mode in the H band and \url{https://www.eso.org/observing/dfo/quality/SPHERE/reports/FULL/trend_report_IRDIS_CLI_DIST_Y_FULL.html} for the classical imaging mode in the Y band.}. A square grid of transparent dots engraved in a layer of black chrome is located in the calibration unit of SPHERE\cite{Wildi2009} hence is common to all SPHERE science instruments. The pitch of the grid is 100.0$\pm$0.5~$\mu$m for a size of the dots of 30~$\mu$m. \rev{Figure~\ref{fig:distmapgrid} shows a map of the distortion vectors.} Distortion grid data are obtained typically once a week for several filter+coronagraph configurations as part of the daytime calibrations. Very few measurements were obtained for the dual-band DB\_Y23 and DB\_J23 filters and are not considered in the analysis. All the data obtained starting from May 2015 were used, after a saturation issue was corrected. Only one technical intervention which could have affected the optical distortion was performed since the beginning of the regular monitoring of the optical distortion. A technical intervention on the cylindrical mirrors occurred on 2015 May 22 UT, very close to the beginning of the monitoring. For a given filter and coronagraph configuration, the temporal evolution of the pixel scale ratio with time shows periodic variations of $\sim$1~yr, with larger values measured during the winter seasons (Fig.~\ref{fig:sphanamorphtime}). The evolution of the bench temperature with time shows variations of similar periodicity but with opposite trends. The evolution of the pixel scale ratio with the temperature of the instrument bench shows an anti-correlation, with larger values measured for lower temperatures\footnote{\rev{We could not look for potential correlations with other parameters related to the ambient conditions because only the bench temperature is recorded in the database of the ESO Quality Control Group.}}. The pixel scale ratio values shown in Table~\ref{tab:anamorphfilt} were computed using a resistant mean procedure rejecting outliers using sigma clipping at 3$\sigma$. The uncertainties are the standard deviation of the measurements obtained over time. The pixel scale ratio does not depend on the use or not of an apodized pupil Lyot coronagraph for a given filter. This is expected because the coronagraph has only a local effect on the distortion measured. The pixel scale ratio has similar values for all broad-band filters and the dual-band DB\_K12 filter, but has a smaller value for the dual-band DB\_H23 filter. For a given configuration, the pixel scale ratio is stable within $\sim$0.3--0.5~mas, within the baseline astrometric requirements (5~mas).

Given the measured uncertainties, the pixel scale ratios measured using on-sky data and the internal distortion grid data are compatible.

\subsection{Pixel scale}
\label{sec:pixscale}

The pixel scale slightly depends on the spectral filter and also on the use of a coronagraph. The main coronagraphs of SPHERE are apodized pupil Lyot coronagraphs and the focal plane masks are deposited on glass plates. We determined that the non-coronagraphic images have smaller pixel scales than the coronagraphic images by a factor of 1.0015. The difference between the pixel scales corresponds to an astrometric uncertainty of $\sim$1.5~mas at 1$''$.

Individual pixel scale measurements are provided in Tables~\ref{tab:pixscalefiltfield}, \ref{tab:pixscalefilt}, and \ref{tab:astrocal}. The uncertainties in Tables~\ref{tab:pixscalefiltfield} and \ref{tab:pixscalefilt} are the standard deviations of the individual measurements in Table~\ref{tab:astrocal}. The uncertainties in Table~\ref{tab:astrocal} are the standard deviations of the measurements on the stellar pairs and are conservative. \rev{The standard way to compute the measurement uncertainty would be to divide the standard deviation by the square root of the number of stars minus 1. However, these measurement uncertainties would be optimistic if applied to calibrate science data not taken close in time to the calibration data because they would not include variations in the instrumental/ambient conditions in the meantime. The observing runs in the SHINE survey typically last a few days, with the astrometric calibration performed once at the beginning of the run.} The uncertainties can vary between measurements on a given stellar cluster field (differences in integration times, observing conditions, S/N, ratio of the point-spread function or PSF width over the S/N). Table~\ref{tab:pixscalefiltfield} compares the statistics on the pixel scale measured in four filters for the 47~Tuc and NGC3603 fields. Table~\ref{tab:pixscalefilt} gives the pixel scale estimates for almost all IRDIS filters (except for the BB\_Y filter, in which no measurements were taken) for the N\_ALC\_YJH\_S apodized pupil Lyot coronagraph. We considered 47 Tuc as the reference field (Sec.~\ref{sec:astrometricfields}). For the filters with no available observations with 47 Tuc, we accounted for pixel scale systematics between the observed field and 47 Tuc by using pixel scale measurements from the two fields obtained with the H2 filter with coronagraph. The specifications were 12.25$\pm$0.01~mas/pix. We note a decreasing trend for the pixel scale with the central wavelength of the filter up to the H band, followed by an increasing trend for longer wavelengths.

\begin{table}[t]
\centering
\caption{SPHERE/IRDIS pixel scales (in mas/pix) measured in different filters for the 47~Tuc and NGC3603 fields.}
\label{tab:pixscalefiltfield}
\begin{tabular}{ccccc}
\hline
Filter & H2 & H3 & K1 & K2 \\
\hline
47~Tuc & 12.250$\pm$0.004 & \rev{12.244}$\pm$0.003 & 12.258$\pm$0.004 & 12.253$\pm$0.003 \\
NGC3603 & 12.245$\pm$0.004 & 12.241$\pm$0.004 & 12.253$\pm$0.004 & 12.249$\pm$0.004 \\
\end{tabular}
\end{table}

\begin{table}[t]
\centering
\caption{Reference value of the SPHERE/IRDIS pixel scale as a function of the filter.}
\label{tab:pixscalefilt}
\begin{tabular}{ccccccccc}
\hline
Filter & Y2 & Y3 & J2 & J3 & H2 & H3 & K1 & K2 \\
\hline
Scale & 12.278 & 12.278 & 12.249 & 12.246 & 12.250 & \rev{12.244} & 12.258 & 12.253 \\
(mas/pix) & $\pm$0.009 & $\pm$0.009 & $\pm$0.009 & $\pm$0.009 & $\pm$0.004 & $\pm$0.003 & $\pm$0.004 & $\pm$0.003 \\
\hline
Filter & BB\_J & BB\_H & BB\_Ks & & & & & \\
\hline
Scale & 12.262 & 12.246 & 12.266 & & & & & \\
(mas/pix) & $\pm$0.009 & $\pm$0.009 & $\pm$0.009 & & & & & \\
\end{tabular}
\end{table}

\begin{figure}[t]
\centering
\includegraphics[width=.49\textwidth]{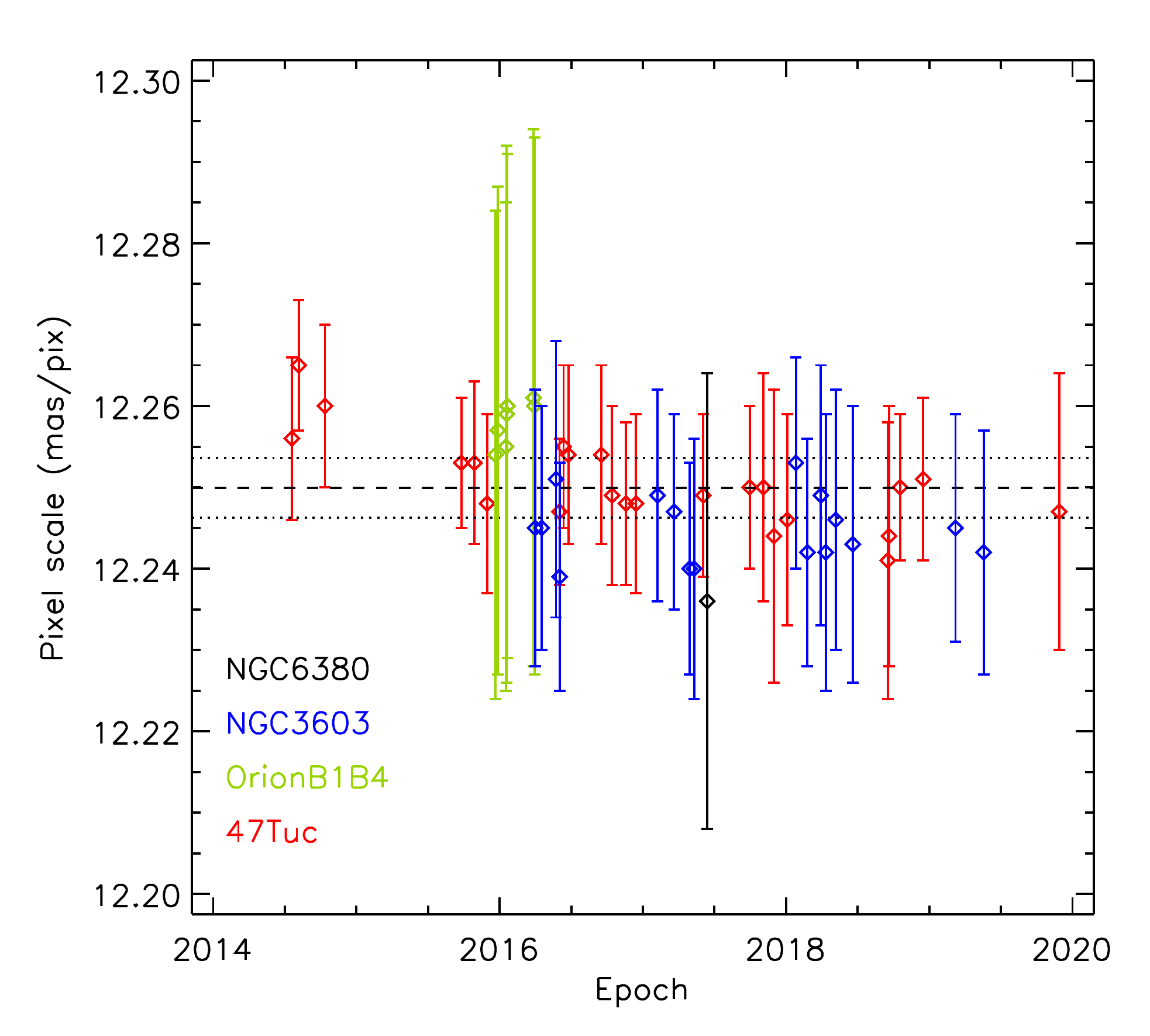}
\caption{Temporal evolution of the pixel scale measured in the SPHERE/IRDIS H2 filter with coronagraph. The dashed line shows the weighted mean of the 47~Tuc measurements obtained starting from 2015 and the dotted lines the 1$\sigma$ interval. \rev{The computation of the uncertainties is explained in Sec.~\ref{sec:pixscale}.}}
\label{fig:pixscale}
\end{figure}

Figure~\ref{fig:pixscale} shows the temporal evolution of the pixel scale for coronagraphic images obtained in the H2 filter since the commissioning (see Table~\ref{tab:astrocal}). The weighted mean and standard deviation of the measurements starting from 2015 are 12.250$\pm$0.004~mas/pixel on 47~Tuc and 12.245$\pm$0.004~mas/pixel for NGC3603. Given the uncertainties, we do not notice a significant systematic between the calibration fields. The standard deviation measured for 47~Tuc translates into an uncertainty at 1$''$ of 0.33~mas, within the baseline astrometric requirements. The weighted mean and standard deviation of the pixel scale measured on 47~Tuc during the commissioning are 12.261$\pm$0.005~mas/pix. Except for measurements obtained during the commissioning, SPHERE has demonstrated a good astrometric stability over five years.

\subsection{North offset correction angle}
\label{sec:northangle}
The correction angle to the North does not depend on the spectral filter nor on the use of a coronagraph. Individual measurements are provided in Table~\ref{tab:astrocal}. The SPHERE images shall be \rev{rotated by the correction angle to the North} to align them with North up. The values are negative, so the rotation is to be performed in the clockwise direction (Fig.~\ref{fig:astrocalpos}). The uncertainties are the standard deviations of the measurements on the stellar pairs and are conservative \rev{(see Sec.~\ref{sec:pixscale})}. For the measurements on the stellar clusters, the error bars can vary from one observation to another because of the sensitivity (integration time, use or not of a coronagraph) and the quality of the images (ratio of the S/N over the PSF width). 

\begin{figure}[t]
\centering
\includegraphics[width=.49\textwidth]{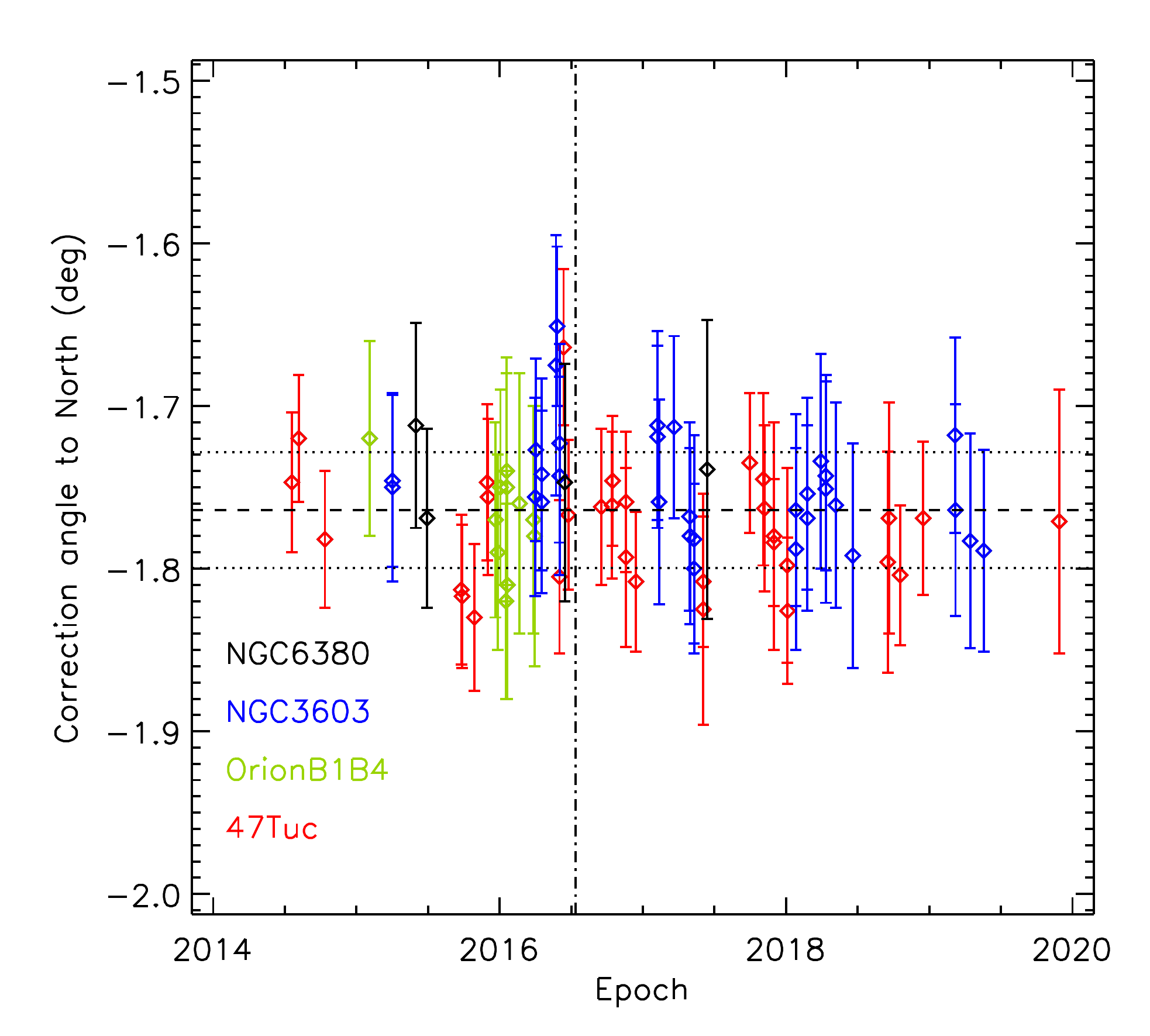}
\caption{Temporal evolution of the correction angle to the North measured with SPHERE/IRDIS data. The dotted-dashed vertical line indicates the epoch when the time reference issue was solved\cite{Maire2016b}. All previous measurements were corrected a posteriori (Sec.~\ref{sec:timerefissue}). The dashed horizontal line shows the weighted mean of all the measurements and the dotted lines the 1$\sigma$ interval. More measurements are shown compared to Fig.~\ref{fig:pixscale} showing the pixel scale because the correction angle to the North does not depend on the filter and coronagraph configuration. \rev{The computation of the uncertainties is explained in Sec.~\ref{sec:pixscale}.}}
\label{fig:tn}
\end{figure}

Figure~\ref{fig:tn} shows the temporal evolution of this parameter. For a given epoch in dual-band imaging mode, we show the value measured for the left-hand filter (H2, K1, J2). We do not see significant variations of this parameter before and after the correction of the time reference issue (\rev{-1.76}$\pm$0.04~deg vs. -1.77$\pm$0.03~deg) and also between the different calibration fields used (-1.77$\pm$0.04~deg for 47~Tuc vs. -1.75$\pm$0.03~deg for NGC3603). The weighted mean and standard deviation for all the measurements are -1.76$\pm$0.04~deg. The standard deviation is within the requirement on the precision for the position angle (0.2$^{\circ}$). It translates into an uncertainty at 1$''$ of 0.70~mas, within the baseline astrometric requirements. Over 5 years of operations, this parameter has been stable.

\subsection{Dependency of the pixel scale and correction angle to the North with the number of stars}

\begin{figure}[t]
\centering
\includegraphics[width=.99\textwidth]{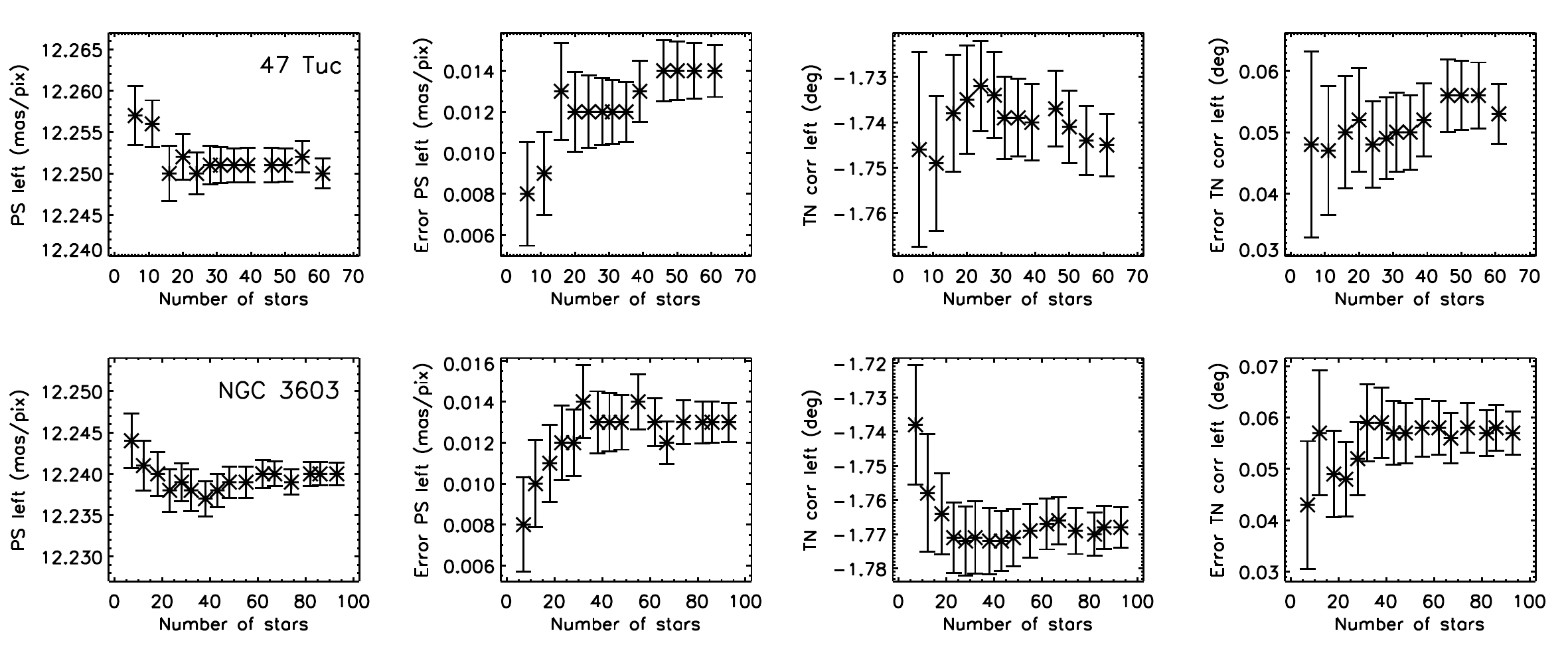}
\caption{Pixel scale \rev{(noted PS)}, standard deviation of the pixel scale, correction angle to the North \rev{(noted TN corr)}, and standard deviation of the correction angle to the North as a function of the number of calibration stars for one dataset acquired for 47~Tuc (top) and NGC3603 (bottom). Measurements in the left-hand SPHERE/IRDIS dual filter were used.
}
\label{fig:evolcalibnstars}
\end{figure}

Part of the variations observed in the pixel scale and North correction angle for the stellar cluster fields could be due to the use of different numbers of stars from one epoch to another, depending on the data quality. Figure~\ref{fig:evolcalibnstars} shows the evolution of the pixel scale, standard deviation of the pixel scale, correction angle to the North, and standard deviation of the correction angle to the North as a function of the number of calibration stars for one dataset acquired for 47~Tuc and NGC3603. For 47~Tuc, most parameters do not vary significantly when more than 16 stars are used\rev{. The correction angle to the North varies even when several tens of stars are used but within the uncertainties. The variations} ($\sim$0.01$^{\circ}$) are smaller than the variations seen over time (0.04$^{\circ}$). For NGC3603, most parameters do not vary significantly when more than 23 stars are used, whereas the standard deviation on the correction angle to the North does not vary significantly when more than 32 stars are used. Thus, the variations in the pixel scale and North correction angle over time in Table~\ref{tab:astrocal} and Figs.~\ref{fig:pixscale} and \ref{fig:tn} cannot be explained by the use of different sets of stars.

\subsection{Analysis of the ESO calibration data}
\label{sec:esocalib}


The pixel scale and correction angle to the North of IRDIS were also monitored in the ESO calibration plan. The same fields as used for the SHINE survey in 47~Tuc, NGC6380, and $\theta^1$ Ori B1--B4 were observed on a monthly basis, mostly without coronagraph \rev{in several IRDIS filters: the narrow-band filters DB\_H23 and DB\_K12 and the broad-band filters BB\_Y, BB\_J, BB\_H, and BB\_Ks}. In addition, we also analyzed some datasets obtained as part of the technical time program. We analyzed the data with the same methods used for the SHINE survey data. \rev{We chose to analyze separately the SHINE and ESO calibration data to assess the impact of not using a coronagraph on the precision achieved for the astrometric calibration.} Our analysis revealed that $\sim$\rev{40\% of the data obtained in the ESO calibration plan} were not suitable for deriving a calibration, mostly because of their lack of depth. \rev{The data obtained on 47~Tuc and NGC6380 are particularly affected by the sensitivity issue, with $\sim$70\% and $\sim$30\% of the data not usable, respectively.} In particular, broad-band observations were obtained with a neutral density filter \rev{to avoid saturation of the AO guide star}. 

\begin{figure}[t]
\centering
\includegraphics[width=.98\textwidth]{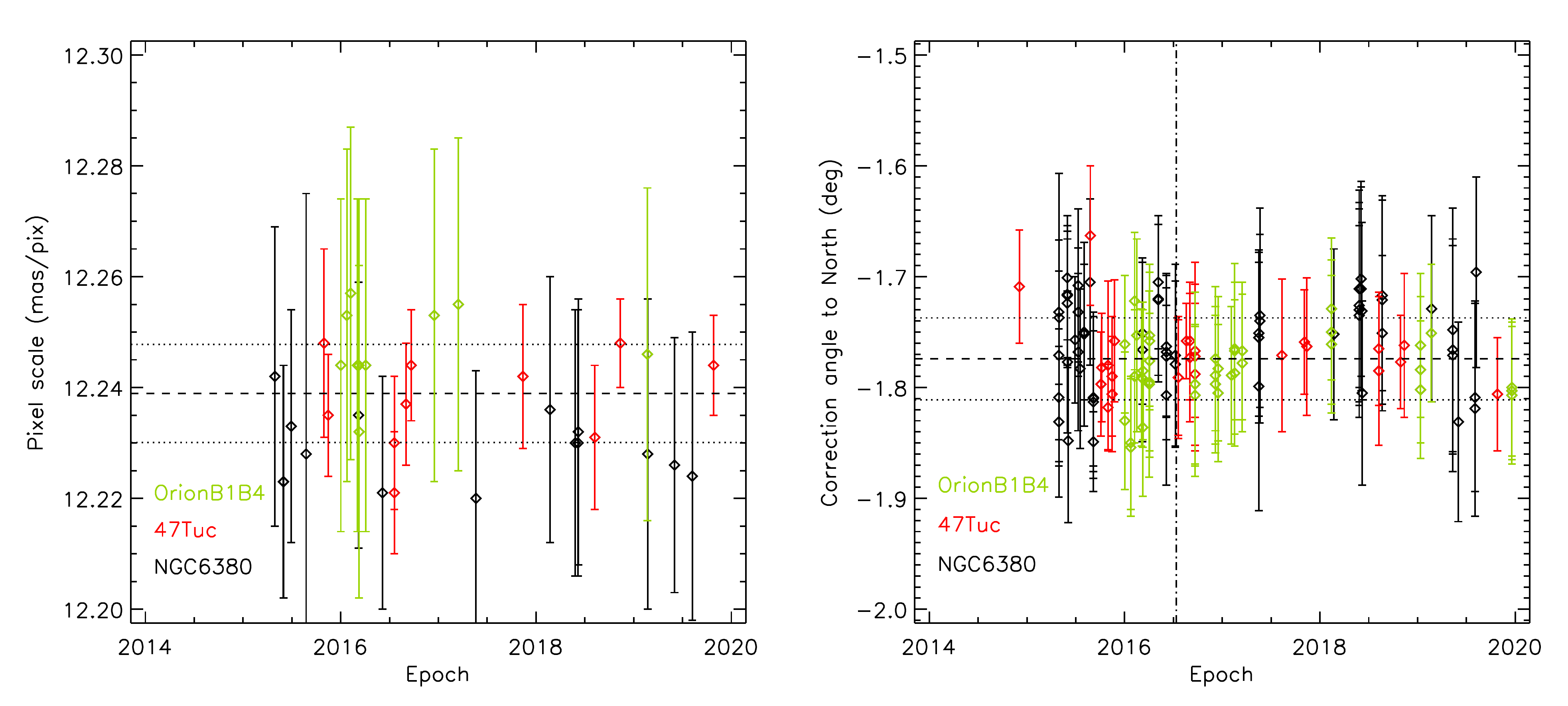}
\caption{Temporal evolution of the pixel scale (left) and correction angle to the North (right) of SPHERE/IRDIS measured with the ESO calibration data. For the pixel scale, only measurements in the H2 filter without a coronagraph are shown. See also the captions of Figs.~\ref{fig:pixscale} and \ref{fig:tn}.}
\label{fig:pstneso}
\end{figure}


Table~\ref{tab:astrocaleso} provides the measurements on the data sets that we could analyze. Figure~\ref{fig:pstneso} shows the individual measurements of the pixel scale in the H2 filter without coronagraph and of the correction angle to the North. The weighted mean and standard deviation of the pixel scale measured on 47~Tuc in the H2 filter without coronagraph and of the correction angle to the North are \rev{12.239}$\pm$0.009~mas/pix and \rev{-1.77$\pm$0.04}~deg, respectively. The uncertainty on the pixel scale is larger than the uncertainty determined for the coronagraphic H2 pixel scale in the SHINE survey data by more than a factor 2. The pixel scale value is smaller compared to the value measured with the SHINE survey data because no coronagraph was used. The value is also consistent with the expected value estimated by dividing the coronagraphic pixel scale measured with the SHINE survey data by the ratio of the coronagraphic and non-coronagraphic pixel scales determined using the SHINE survey data (Sec.~\ref{sec:pixscale}). The correction angle to the North agrees with the value determined using the SHINE survey data, with \rev{a similar uncertainty}.

\subsection{\rev{Correction angle for the pupil zeropoint angle in pupil-tracking mode}}
\label{sec:ptftoffset}

\begin{figure}[t]
\centering
\includegraphics[width=.88\textwidth]{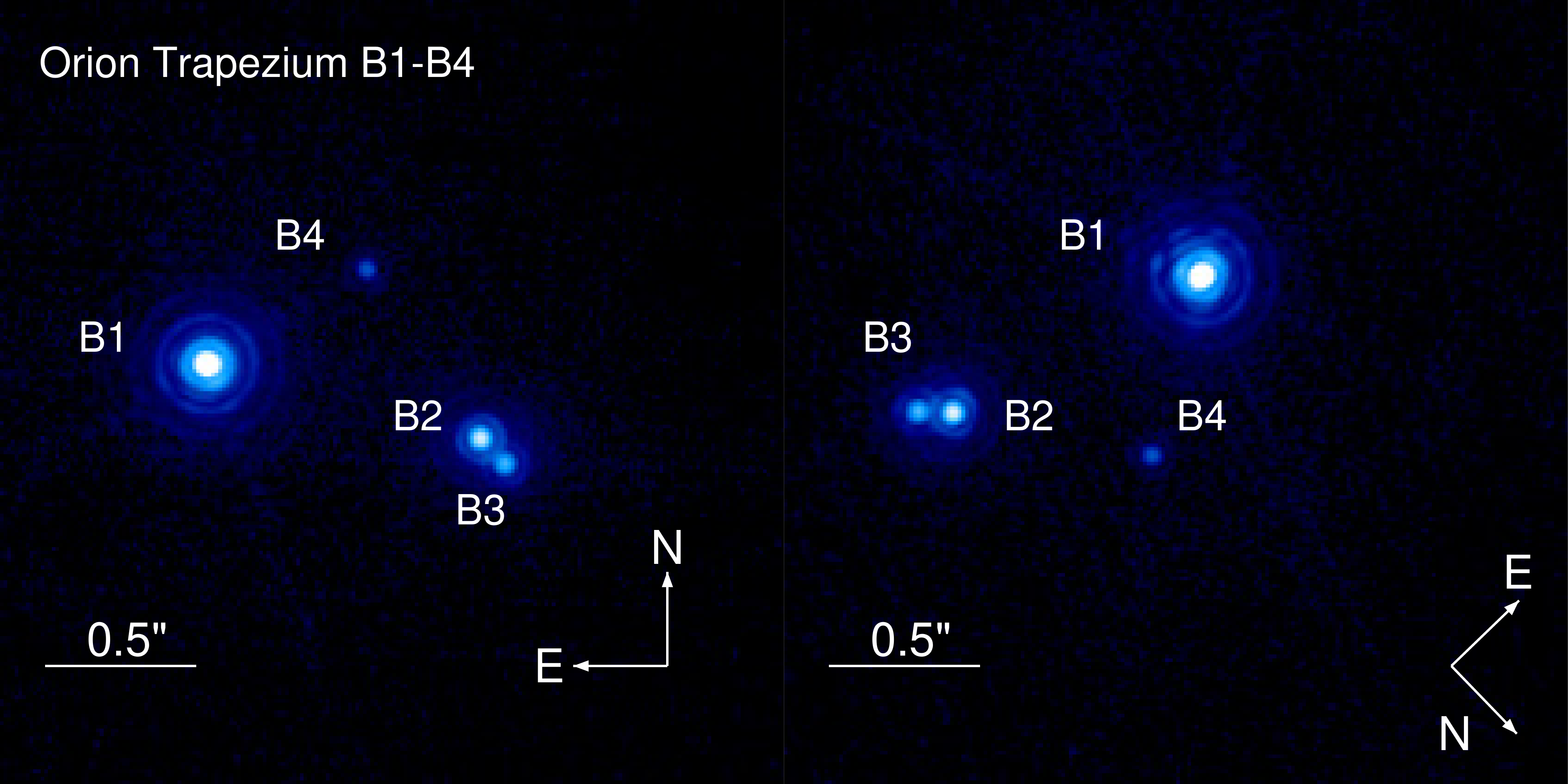}
\caption{SPHERE/IRDIS images of the $\theta^1$ Ori Trapezium B1–B4 taken in field-tracking mode (left) and in pupil-tracking mode (right) for the same position angle of the image derotator.
}
\label{fig:ptftangle}
\end{figure}

To align North up and East to the left, SPHERE images obtained in pupil-stabilized mode shall also be derotated from the zeropoint angle of the instrument pupil in this mode (Fig.~\ref{fig:ptftangle}). We found that \rev{the correction angle for the pupil zeropoint} is stable using data taken in the first two years of SPHERE operations. We used data of several fields observed consecutively in field-stabilized mode and pupil-stabilized mode (with similar pointing parameters). The measurements are shown in Table~\ref{tab:ptftoffset}. For binary systems and $\theta^1$ Ori B1--B4, the mean value is computed using measurements obtained from good individual frames in the sequences. For 47~Tuc, the mean value is computed using all the stars present in both combined images using sigma clipping. The uncertainties are the standard deviations of the measurements. The statistics of the measurements gives a weighted mean value of \rev{136.00$\pm$0.03$^{\circ}$}. Thus, pupil-tracking images shall be \rev{rotated in the counter-clockwise direction}. The standard deviation is within the requirement on the precision for the position angle and translates into an uncertainty at 1$''$ of 0.52~mas. No new measurements were taken after August 2016 either in the SHINE survey or in the ESO calibration plan.

\renewcommand{\arraystretch}{1.}
\begin{table}[t]
\centering
\caption{Measurements of the \rev{correction angle for the pupil zeropoint angle in pupil-tracking mode}.}
\label{tab:ptftoffset}
\begin{tabular}{ccc}
\hline
Date & Target & Value \\
 & & (deg) \\
\hline
2014-07-18 & 47~Tuc & 135.987$\pm$0.042 \\
2015-02-03 & $\theta^1$ Ori B1--B4 & 135.967$\pm$0.044 \\
2015-02-04 & $\theta^1$ Ori B1--B4 &136.046$\pm$0.100 \\
2015-11-29 & 47~Tuc & 136.023$\pm$0.050 \\
2015-12-01 & 47~Tuc & 136.009$\pm$0.065 \\
2016-06-23 & PZ Tel & 136.011$\pm$0.034 \\
2016-07-14 & HD~130940 & 135.991$\pm$0.035 \\
2016-07-14 & HD~130940 & 135.964$\pm$0.035 \\
2016-08-02 & HD~130940 & 136.022$\pm$0.024 \\
\hline
\hline
Weighted mean & & 136.00$\pm$0.03 \\
\end{tabular}
\end{table}

\begin{table}[t]
\centering
\caption{Astrometric error budgets estimated from the SPHERE calibration for a separation of 1$''$. \rev{All the individual error terms scale linearly with the separation, so that the error budget can be computed for other separations.}}
\label{tab:summaryastrocal}
\begin{tabular}{cccccc}
\hline
Configuration & Distortion & Pixel scale & North angle & Pupil ZP & Total \\
& & & & angle & \\
\hline
SPHERE (mas) & 0.4 & 0.33 & 0.70 & 0.52 & 1.01 \\
ELT scaled, H band (mas) & 0.08 & 0.07 & 0.15 & 0.11 & 0.21 \\
ELT scaled, L$^{\prime}$ band (mas) & 0.19 & 0.16 & 0.34 & 0.25 & 0.49 \\
\end{tabular}
\end{table}

\subsection{Summary}
Table~\ref{tab:summaryastrocal} summarizes the astrometric error budget associated to the uncertainties in the SPHERE calibration. \rev{It is computed for a separation of 1$''$ but can be computed for other separations because all the individual error terms scale linearly with the separation.} It provides accuracy limits to which astrometry-relevant instrument properties could be calibrated. To reach this level of calibration, an optimized calibration strategy is needed, which needs to reflect the instrumental stability of the parameters, listed in Table~\ref{tab:summaryastrocal}. However, we note that in practice, in the case of faint companions, the achievable S/N limits the centroid precision to values larger than the astrometric performance listed in Table~\ref{tab:summaryastrocal}, so that the uncertainties in the calibration do not add significantly to the total error budget. Thus, the need for optimal astrometric calibration could be relaxed. Here, we assume instead the case of companions detected at high S/N so that the measurement uncertainties are smaller than the calibration uncertainty (S/N larger than the ratio of the PSF width over the calibration uncertainty, assuming that the astrometric uncertainty is given by the ratio of the PSF width over the S/N)\cite{Lindegren1978}.

The positional uncertainty due to the astrometric calibration of SPHERE amounts to 1.01~mas at 1$''$, which is within the baseline astrometric requirement and at the level of the goal requirement. The uncertainties on the correction angle to the North and of the pupil zeropoint angle in pupil-tracking mode are the main contributors to the error budget. The uncertainties on the pixel scale and on the correction angle to the North were included in the computation of the uncertainties on the position of the companions provided by the SPHERE Data Center but not the uncertainty in the correction angle of the pupil zeropoint angle in pupil-tracking mode and, for IFS, the uncertainty in the angle offset with respect to the IRDIS FoV. This was corrected in July 2020.

The next step for exoplanet imaging will be the use of extremely large telescopes, in particular the ELT. Due to the combination of increased angular resolution and collecting aperture, diffraction-limited ELT observations will at the same time access smaller angular separations, and achieve higher astrometric precision at angular separations accessible to 8m-class imagers. Assuming that all the astrometric uncertainties listed in Table~\ref{tab:summaryastrocal} scale as $\lambda/D$, we also estimated a calibration error budget for a high-contrast imaging instrument on the ELT operating in the H band, which is covered by the Multi-adaptive optics Imaging Camera for Deep Observations (MICADO)\cite{Davies2018} and the High Angular Resolution Monolithic Optical and Near-infrared Integral field spectrograph (HARMONI)\cite{Thatte2016}, and in the L$^{\prime}$ band, which is covered by the Mid-infrared ELT Imager and Spectrograph (METIS)\cite{Brandl2018}. Under this assumption and assuming that the measurement uncertainties in the position of detected companions are small compared to the calibration uncertainty (i.e., companions detected with S/N larger than the ratio of the PSF width over the calibration uncertainty), we expect that sub-mas precisions should be achieved with exoplanet imaging instruments on the ELT.

\section{Lessons learned for the next generation of exoplanet imaging instruments}
\label{sec:lessons}

\subsection{Instrument stability}
\label{sec:instrustability}
Instruments aiming to high-precision astrometry over time should be stable against temperature variations, gravity flexures, and pupil rotation. SPHERE operates on a Nasmyth focus, which provides stability against gravity flexures and temperature variations but requires to compensate precisely for the pupil rotation in both field-tracking and pupil-tracking modes. In comparison, the dedicated exoplanet imaging instrument Gemini/GPI was located on a Cassegrain focus, which provides stability against pupil rotation but is sensitive to temperature variations and gravity flexures. The ELT instruments will operate on Nasmyth focii, so a precise control of the pupil rotation will be mandatory for high-precision astrometry.

Also important in the SPHERE stability is the absence of significant technical interventions on the instrument since it was made available to the community. In comparison, the first-generation exoplanet imaging instrument VLT/NaCo received regular technical interventions to implement new observing modes or fix issues and was moved to another Unit Telescope (UT). GPI was removed yearly from the Gemini telescope due to telescope shutdowns. This may cause the slight trend of increasing north offset angle of $\sim$0.2$^{\circ}$ found over 6 years of data\cite{DeRosa2020}.

\subsection{Frame registration}
High-contrast imaging observations are commonly performed in pupil-tracking mode, which allows for fixing the aberrations in the images due to the instrument defects conjugated to the telescope pupil. As a result, the FoV rotates around the on-axis star. A good frame registration is critical for high-precision astrometry and for maximizing the performance of high-contrast imaging algorithms. This maximizes the S/N of the measured companions and minimizes measurement uncertainties and biases on their position.

The uncertainties in the determination of the location of the star (either saturated or behind a coronagraph) was a major limitation to the astrometric precision of the first generation of exoplanet imaging instruments. To meet their astrometric requirements, the dedicated exoplanet imaging instruments SPHERE and GPI included a specific strategy to monitor the location of the star as close as possible in time to the science observations.

Producing calibration spots using the AO deformable mirror as done for SPHERE presents two advantages compared to using a diffraction grid as used for the GPI instrument. First, the brightness of the spots can be tuned by adjusting the amplitude of the periodic modulation to ensure a good S/N for their detection when observing faint stars or when the observing conditions are poor (higher noise from the AO halo). Then, the orientation of the spots can be modified to avoid that a companion having a similar angular separation as the spots crosses one of them due to the field rotation. Another lesson drawn from SPHERE observations is that real-time monitoring of the location of the star during the science observations is required to achieve the best astrometric precision because of instabilities in the star centering (Sec.~\ref{sec:starceninstability}). \rev{One drawback of the use of the calibration spots is a slight decrease of the AO performance, with slightly smaller Strehl ratios (ratio of the peaks of the measured and theoretical non-coronagraphic PSF).}

Implementing the SPHERE strategy for the ELT instruments may be more difficult because the AO deformable mirror for the ELT is part of the telescope, so its design is not under the responsibility of the instrument consortia. Instead, implementing the GPI strategy of a diffraction grid should be easier, but with the limitations mentioned above.  

\subsection{Image derotation}
A good image derotation is mandatory to realign the images of the companions right after the subtraction of the stellar residuals with high-contrast imaging algorithms. A poor re-alignment will result in a poorer S/N of the measured companions and larger measurement uncertainties on their position as well as biases for the position angle.

Image derotation requires to compute the parallactic angles of each individual image. This can be done using the stellar coordinates at the observing epoch and the timestamps of the individual images. The stellar coordinates at the observing epoch are not always provided in the data headers but instead the stellar coordinates at a reference epoch such as J2000. In this case, the coordinates need to be corrected for the precession. Computing the individual timestamps requires a good knowledge of how the images are exactly recorded (overheads), because the data headers typically provide only two timestamps, when the data recording is started and when the data file is written on disk. Good communication between instrument and survey teams is important to ensure that the information is available at the start of the scientific operations of an instrument.

\subsection{Astrometric calibration}
To achieve high-precision astrometry with SPHERE, the monitoring of several parameters is required: optical distortion, pixel scale, correction angle to the North, and correction angle of the pupil zeropoint in pupil-tracking mode. Given the large number of observing modes (filter, coronagraph) and that the values of some astrometric parameters depend on the observing mode, a careful optimization of the calibration plan was mandatory to minimize the use of the telescope time during the night. 

SPHERE has displayed astrometric stability over 5 years of operations. This could be demonstrated thanks to a homogeneous calibration strategy and the stability of the instrument. Astrometric stability is critical given the long timescales needed to constrain the orbital parameters of directly-imaged exoplanets. It also allows for saving telescope time because it relaxes the need to take nighttime astrometric data close to the science observations for a precise calibration. Astrometric stability will be even more relevant for the ELT, because of the high pressure expected on this facility.

The measurement uncertainty on the correction angle of the pupil zeropoint in pupil-tracking mode is currently the second main limitation to the SPHERE astrometric precision. This parameter was monitored only during the first two years of operations. It was not monitored as part of the ESO calibration plan due to software issues. Work is ongoing with the ESO Staff to monitor this parameter in the ESO calibration plan. Although we expect this parameter to be stable, new data may help to refine its estimate. Most data used for the current analysis were corrected a posteriori for the time reference issue (Sec.~\ref{sec:timerefissue}), so the precision of the measurements may be limited by uncertainties in the correction. Another limitation to the measurement precision of this parameter could be the derotator backlash (Sec.~\ref{sec:derotbacklash}).

Another lesson learned from SPHERE is that the astrometric calibration of coronagraphic images when using stellar fields with low densities (binaries, multiples) requires to shift the guide star out of the coronagraph. This comes for free if the photometric calibration is based on images of the star observed out of the coronagraph. This was the case for SPHERE but not for the GPI instrument. Due to its small FoV, GPI could only observe stellar binaries and multiples as calibrators. Obtaining calibration data in coronagraphic mode was only feasible for the stellar multiple $\theta^1$ Ori B1--B4. However, the calibration suffers from large uncertainties because only the two close stars B2--B3 could be measured (separation $\sim$0.12$''$).


Most astrometric data taken as part of the SHINE survey were obtained with a coronagraph. Astrometric data without a coronagraph were taken monthly as part of the ESO calibration plan. However, our analysis showed that $\sim$\rev{30--70\% of the data obtained on the stellar clusters NGC6380 and 47~Tuc} are not suitable for deriving a calibration, because of their lack of depth (short exposure time). Since regular astrometric observations are important to monitor changes in the calibration and that no more SHINE survey data will be available after September 2021 due to the survey completion, work is ongoing with the ESO staff to increase the exposure time of the observations.

In contrast to SPHERE, the astrometric calibration for the VLT/NaCo instrument was heterogeneous, irregular, and mostly left to the observing teams\cite{Chauvin2012,Ginski2014,Plewa2015}. Accounting also for the technical interventions on the instrument (Sec.~\ref{sec:instrustability}), this resulted in poor astrometric stability, making the use of the data for high-precision astrometry more difficult. The limitations encountered with NaCo were taken into account in the design of SPHERE (gravity invariance, dedicated coronagraph tip/tilt loop).

\subsection{Exoplanet imaging with the ELT}
Exoplanet imaging is part of the science cases of the first three ELT instruments: MICADO, HARMONI, and METIS. MICADO is the only instrument for which the design is driven by astrometric requirements for the observation of the Galactic Center and wide stellar fields\cite{Davies2018}. The requirements in regular imaging are 50~$\mu$as, with a goal of 20~$\mu$as\cite{Pott2018}. This will be achieved thanks to a dedicated astrometric stability by design and built-in calibration strategy. However, high-precision astrometry in exoplanet imaging requires a dedicated strategy, as described in this paper using SPHERE as an example.

Because MICADO, HARMONI, and METIS will study different types of exoplanets, an astrometric strategy should be implemented for each instrument. While MICADO (0.8--2.4~$\mu$m) and HARMONI (0.47--2.45~$\mu$m) will overlap for near-IR observations, the spectroimaging capabilities of HARMONI will allow for reaching deeper constrasts hence fainter or less massive young giant exoplanets\cite{Houlle2021}. METIS (3--13~$\mu$m) will cover thermal IR wavelengths and will study more mature exoplanets and less massive exoplanets down to rocky planets\cite{Carlomagno2020}.

The first limiting factor to the astrometric precision of the ELT instruments will be the ELT opto-mechanical stability. The ELT will experience temporal variations of plate scale and field orientation. A control loop will correct for drifts of the M2 mirror with respect to the M1 mirror due to wind disturbances and gravity flexures and recollimate it every 5 minutes\cite{Mueller2014}. The drift is expected to produce variations of plate scale up to $\sim$5~mas/arcmin over 5 minutes, i.e. $\sim$0.08~mas at 1$''$\cite{Rodeghiero2018}. To reach its astrometric requirements, the current strategy for MICADO is to use one of the deformable mirrors of the Multi-conjugate Adaptive Optics RelaY (MAORY) to control the variations of plate scale in between two corrections of the control loop. However, this strategy is not applicable for single conjugated AO, which is the most commonly-used AO mode for exoplanet imaging. Variations of field orientation will be produced by small relative tilts between the plane and adaptive M4 and M5 mirrors\cite{Rodeghiero2018}. The field rotation due to the AO tip-tilt correction of the M5 mirror cannot be avoided, posing some limitations to the accuracy of the centroiding of the PSF in the outer parts of the FoV. The effect is expected to translate into an elongation of the PSF of 1.5~mas in H band and 3~mas in K band at a radius of 30$''$. Exoplanet imaging uses smaller FoVs, so such observations will be less affected by the effect, 0.05~mas in H band and 0.1~mas in K band at 1$''$. Estimating the amplitude of the effect for METIS observations in the L$^{\prime}$ band assuming a linear extrapolation gives $\sim$5~mas at a radius of 30$''$ and $\sim$0.17~mas at 1$''$. The estimates of the plate scale and field orientation variations expected for the ELT are within the astrometric error budget for ELT instruments derived in Table~\ref{tab:summaryastrocal}. Tests during the ELT commissioning will be critical to verify that the actual magnitudes of the plate scale variations and of the field rotations agree with the expected values mentioned above.

The differences in instrument concepts (e.g., FoV, sensitivity, observing modes) will affect the astrometric strategy. HARMONI will cover small FoVs in its high angular resolution modes compared to MICADO and METIS ($\lesssim$4$''$ against $\sim$10$''$). Finding astrometric fields with a large number of stars in such small FoVs may be challenging and prevent on-sky measurements of the optical distortion. A possible strategy for calibrating HARMONI could be to use parallel observations of a same astrometric field with an instrument with a good calibration or to combine a large number of observations of asteroids with well-constrained orbits to construct a distortion map. The first strategy was employed for calibrating the GPI instrument using observations obtained close in time with the Keck/NIRC2 camera\cite{DeRosa2020}. For METIS, the faintness of the stars at thermal IR wavelengths compared to the background noise will likely restrict usable astrometric fields to fields with bright stars. Also in this case, measuring on sky the optical distortion may be challenging. Using internal distortion grids to measure the optical distortion will still be feasible, but the distortion from the ELT optics will not be measured. METIS will include an imager and long-slit spectrograph in the L, M, and N bands and an integral field spectrometer (IFS) at high spectral resolutions ($\sim$100\,000) with a FoV of $1''\times0.5''$ covering only the L and M bands. Parallel observations are foreseen between the IFS and the imager in the L and M bands, so that an astrometric calibration of the IFS relative to the imager could be done easily using only internal distortion grid data to measure the pixel scale ratio and a potential angle offset between the FoVs of the instruments without further nighttime calibration data needed. We employed such a strategy to calibrate the IFS of SPHERE on the IRDIS camera\cite{Maire2016b}.

Given the high observing pressure expected for the ELT and the large number of observing modes of the instruments, a careful optimisation of the astrometric calibration plan will be required to minimize the use of nighttime observations. Optimizing the instrument design for stability and maximizing the use of the distortion grid data obtained in daytime will be critical in this respect. Deriving an error budget will be key to identify if one specific term dominates compared to the others. If this is the case, the calibration plan could be balanced out to reduce the uncertainty in the limiting parameter while relaxing the constraints on the other parameters (e.g., taking more frequent but shallower observations). Nighttime observations should be used to derive absolute measurements and/or to calibrate the observing mode that will be most likely used, whereas distortion grid data should be used to derive relative measurements and/or to calibrate the other observing modes on a reference observing mode. For instance, the offset angle between pupil-tracking and field-tracking observations could be measured with a distortion grid if located upstream of the image derotator. 

Gaia and the James Webb Space Telescope (JWST) will provide reference fields with precise and accurate positions for astrometric calibration of future instruments on ground-based telescopes. However, the poor angular resolution of Gaia mainly limits astrometric comparisons with instruments on 8--10~m telescopes to widely-separated binaries ($\gtrsim$5$''$). It will also be a limitation for comparisons with instruments on ELT telescopes. JWST will provide higher angular resolutions than Gaia and will observe down to 0.6~$\mu$m, making astrometric comparisons with the ELT instruments easier. The JWST astrometric calibration will be based on observations of a field of 5$'\times$5$'$ area with a relatively homogeneous distribution of stars in the Large Magellanic Cloud\cite{Diaz-Miller2007}$^,$\footnote{\url{https://jwst-docs.stsci.edu/data-processing-and-calibration-files/absolute-astrometric-calibration}}. The catalog of stellar positions was derived from HST observations with positional accuracies of $\sim$1~mas. In addition, a larger field in the Large Magellanic Cloud observed with the VLT instrument High Acuity Wide field K-band Imager (HAWK-I) including the HST field and covering all the arrays of the JWST instruments Near Infrared Camera (NIRCam) and Fine Guidance Sensor (FGS) at once will be used\cite{Sahlmann2019,Anderson2021}. These calibration fields will allow for deriving a distortion solution for all the JWST instruments to an uncertainty smaller than 5~mas rms along each detector axis. The absolute astrometric reference frame of JWST will be referenced to the Gaia reference frame. The NIRCam calibration field would be suitable to calibrate observations with the ELT and the Giant Magellan Telescope given their location in the southern hemisphere. For cross-checks with the astrometry measured with northern observatories, another calibration field would be needed. Another instrument of interest for the astrometric calibration of ground-based exoplanet imaging instruments would be the Nancy Grace Roman Space Telescope (formerly WFIRST)\cite{WFIRSTAWG2019}.

To estimate and calibrate potential remaining astrometric systematic uncertainties between ELT and VLT instruments, parallel observations will be critical. This was done for the VLT instruments NaCo and SPHERE though NaCo's opto-mechanics was heavily gravity-dependent hence this cross-instrument calibration was only valid within a reduced time frame. This is done now between SPHERE and the interferometric instrument GRAVITY. SPHERE is expected to remain operational until the first years of ELT operations, hence could be used as a reference for the ELT instruments. Such relative measurements would not require an astrometric field with catalog positions with a good absolute calibration. The astrometric field could be selected to be observable from both southern and northern hemispheres to enable homogeneous comparisons between a large number of telescopes. With astrometric precisions better by a factor of $\sim$30 compared to SPHERE, GRAVITY could also be used to test and validate the absolute astrometric calibration of coronagraphic instruments, thanks to the absolute calibration provided by its internal metrology system\cite{Lacour2014}.

\section{Conclusions}


We described in this paper the astrometric strategy and a 5-yr analysis of the astrometric calibration of the SPHERE instrument, the first instrument dedicated to exoplanet imaging at ESO/VLT. The astrometric strategy of SPHERE relies on an observing procedure for a precise determination of the star location behind the coronagraph, an accurate determination of the instrument overheads and metrology, and regular observations of the same fields in stellar clusters for the astrometric calibration. We solved several issues encountered in the course of the on-sky operations. A technical intervention solved the time reference issue and we implemented a correction for the data obtained in the first two years of operations. We revised our strategy for the orbital monitoring of exoplanets and brown dwarfs to use simultaneously with the science observations the calibration spots to monitor the location of the star and minimize stellar centering uncertainties due to instabilities during the sequence. Using the astrometric data obtained during the SHINE survey, we showed that the optical distortion, pixel scale, correction angle to the North, and correction angle of the pupil zeropoint in pupil-tracking mode are stable within a combined error budget of 1~mas for a separation of 1$''$. This is well within the 5-mas baseline requirement for the astrometric precision and matches the goal requirement of 1~mas. The uncertainties on the correction angle to the North (0.7~mas) and on the correction angle of the pupil zeropoint in pupil-tracking mode (0.5~mas) are the main contributors to the error budget. The homogeneous and stable astrometric calibration of SPHERE has facilitated high-precision studies by its users since its start of operation in 2014 by reducing the telescope overheads for nighttime calibration. We also found that $\sim$\rev{30--70\% of the monthly astrometric data taken in 2015--2019 on the stellar clusters NGC6380 and 47~Tuc} as part of the Observatory calibration plan have a suboptimal quality for precise calibrations\rev{, because of their lack of sensitivity}. Work is ongoing with the Observatory Staff to improve the setup of the observations. 

SPHERE being the first instrument dedicated to exoplanet imaging on an ESO telescope, the lessons learned from its astrometric analysis are valuable to optimize the strategy of the exoplanet imaging modes of the ELT instruments MICADO, HARMONI, and METIS. Assuming that all the components of the SPHERE astrometric calibration error budget scale as $\lambda/D$, we estimated that the ELT instruments in coronagraphic imaging mode could achieve astrometric precisions at a separation of 1$''$ of $\sim$0.2~mas in the H band and $\sim$0.5~mas in the L$^{\prime}$ band for companions detected at high S/N (such that the ratio of the PSF width over the S/N is smaller than the calibration uncertainty). High-precision astrometry imposes constraints on various aspects of the design of high-contrast imaging instruments: a large FoV, a good sensitivity, the calibration of the star location behind the coronagraph for precise frame registration, the precision of the image derotator, the astrometric calibration of coronagraphic images when few stars are available in the FoV, and a precise knowledge of the overheads and metrology in the data recording. It also requires an optimization of the calibration plan to monitor all the needed astrometric parameters in the main observing modes while minimizing the needs for nighttime observations. Optimizing the instrument design for stability and/or maximizing the use of the daytime calibration data will be critical. The calibration plan can also be balanced out to meet a top-level astrometric requirement and provide long-term monitoring if there is a dominant error term in the astrometric error budget. The opto-mechanical stability of the telescope may also be the major limitation to the astrometric precision. As a result, the final astrometric precision may depend on the exposure time or length of the observation. In this case, astrometric error budgets should also include timescale requirements.

To maximize the scientific return of future exoplanet imaging instruments for high-precision astrometry, we recommend that clear astrometric requirements (with timescale requirements if applicable) should be established so that they can be used to optimize the instrument design and observing and calibration procedures and that the calibration plan should be optimized to maximize the use of internal distortion grid data to measure relative astrometric parameters so that the use of nighttime observations can be reduced to the direct calibration of the observing modes that will be most likely used.

\appendix

\section{Individual measurements of pixel scale and correction angle to the North}

Table~\ref{tab:astrocal} provides the individual measurements of pixel scale and correction angle to the North derived with the SPHERE/IRDIS data obtained in the SHINE GTO survey. Table~\ref{tab:astrocaleso} provides the measurements derived with the SPHERE/IRDIS data obtained as part of the ESO Paranal calibration program.

\renewcommand{\arraystretch}{.5}
\begin{longtable}{cccr@{}lc}
\label{tab:astrocal}
UT Date & Field & Filter & \multicolumn{2}{c}{Pixel scale} & North correction angle \\
& & & \multicolumn{2}{c}{(mas\,px$^{-1}$)} & ($^{\circ}$) \\
\hline
2014-07-18 & 47Tuc & H2 &  12.256$\pm$&0.010 &  -1.747$\pm$0.043 \\
2014-07-18 & 47Tuc & H3 &  12.251$\pm$&0.010 &  -1.760$\pm$0.048 \\
2014-08-05 & 47Tuc & H2 &  12.265$\pm$&0.008 &  -1.720$\pm$0.039 \\
2014-08-05 & 47Tuc & H3 &  12.259$\pm$&0.009 &  -1.735$\pm$0.045 \\
2014-10-11 & 47Tuc & H2 &  12.260$\pm$&0.010 &  -1.782$\pm$0.042 \\
2014-10-11 & 47Tuc & H3 &  12.254$\pm$&0.010 &  -1.795$\pm$0.044 \\
2015-02-03 & $\theta^1$ Ori B1-B4 & H2 &  12.257$\pm$&0.030$^{b}$ &  -1.720$\pm$0.060 \\
2015-02-03 & $\theta^1$ Ori B1-B4 & H3 &  12.252$\pm$&0.030$^{b}$ &  -1.770$\pm$0.060 \\
2015-03-31 & NGC3603 & BB\_Ks &  12.246$\pm$&0.013$^{b}$ &  -1.750$\pm$0.058 \\
2015-03-31 & NGC3603 & BB\_J &  12.242$\pm$&0.011$^{b}$ &  -1.746$\pm$0.053 \\
2015-05-30 & NGC6380 & H2 &  12.220$\pm$&0.020$^{b}$ &  -1.712$\pm$0.063 \\
2015-05-30 & NGC6380 & H3 &  12.217$\pm$&0.020$^{b}$ &  -1.722$\pm$0.063 \\
2015-06-28 & NGC6380 & H2 &  12.232$\pm$&0.021$^{b}$ &  -1.769$\pm$0.055 \\
2015-06-28 & NGC6380 & H3 &  12.230$\pm$&0.023$^{b}$ &  -1.777$\pm$0.056 \\
2015-09-24 & 47Tuc & H2 &  12.253$\pm$&0.008 &  -1.813$\pm$0.046 \\
2015-09-24 & 47Tuc & H3 &  12.247$\pm$&0.010 &  -1.840$\pm$0.043 \\
2015-09-26 & 47Tuc & K1 &  12.258$\pm$&0.010 &  -1.817$\pm$0.044 \\
2015-09-26 & 47Tuc & K2 &  12.248$\pm$&0.016 &  -1.851$\pm$0.068 \\
\rev{2015-10-27} & 47Tuc & H2 &  12.253$\pm$&0.010 &  -1.830$\pm$0.045 \\
\rev{2015-10-27} & 47Tuc & H3 &  12.246$\pm$&0.009 &  -1.844$\pm$0.040 \\
2015-11-29 & 47Tuc & H2 &  12.248$\pm$&0.011 &  -1.747$\pm$0.048 \\
2015-11-29 & 47Tuc & H3 &  12.244$\pm$&0.011 &  -1.759$\pm$0.051 \\
2015-12-01 & 47Tuc & K1 &  12.254$\pm$&0.011 &  -1.756$\pm$0.048 \\
2015-12-01 & 47Tuc & K2 &  12.251$\pm$&0.015 &  -1.782$\pm$0.064 \\
2015-12-20 & $\theta^1$ Ori B1-B4 & H2 &  12.254$\pm$&0.030 &  -1.770$\pm$0.060 \\
2015-12-20 & $\theta^1$ Ori B1-B4 & H3 &  12.251$\pm$&0.030 &  -1.820$\pm$0.060 \\
2015-12-26 & $\theta^1$ Ori B1-B4 & H2 &  12.257$\pm$&0.030 &  -1.790$\pm$0.060 \\
2015-12-26 & $\theta^1$ Ori B1-B4 & H3 &  12.251$\pm$&0.030 &  -1.830$\pm$0.060 \\
2016-01-02 & $\theta^1$ Ori B1-B4 & K1 &  12.272$\pm$&0.030 &  -1.750$\pm$0.060 \\
2016-01-02 & $\theta^1$ Ori B1-B4 & K2 &  12.267$\pm$&0.030 &  -1.770$\pm$0.060 \\
2016-01-16 & $\theta^1$ Ori B1-B4 & H2 &  12.255$\pm$&0.030 &  -1.820$\pm$0.060 \\
2016-01-16 & $\theta^1$ Ori B1-B4 & H3 &  12.252$\pm$&0.031 &  -1.860$\pm$0.070 \\
2016-01-18 & $\theta^1$ Ori B1-B4 & H2 &  12.259$\pm$&0.033 &  -1.740$\pm$0.070 \\
2016-01-18 & $\theta^1$ Ori B1-B4 & H3 &  12.256$\pm$&0.035 &  -1.750$\pm$0.080 \\
2016-01-18 & $\theta^1$ Ori B1-B4 & K1 &  12.271$\pm$&0.032 &  -1.750$\pm$0.070 \\
2016-01-18 & $\theta^1$ Ori B1-B4 & K2 &  12.268$\pm$&0.033 &  -1.750$\pm$0.080 \\
2016-01-20 & $\theta^1$ Ori B1-B4 & H2 &  12.260$\pm$&0.031 &  -1.810$\pm$0.070 \\
2016-01-20 & $\theta^1$ Ori B1-B4 & H3 &  12.258$\pm$&0.033 &  -1.830$\pm$0.080 \\
2016-02-20 & $\theta^1$ Ori B1-B4 & K1 &  12.274$\pm$&0.031 &  -1.760$\pm$0.080 \\
2016-02-20 & $\theta^1$ Ori B1-B4 & K2 &  12.270$\pm$&0.032 &  -1.770$\pm$0.100 \\
2016-03-26 & $\theta^1$ Ori B1-B4 & H2 &  12.261$\pm$&0.033 &  -1.770$\pm$0.070 \\
2016-03-26 & $\theta^1$ Ori B1-B4 & H3 &  12.259$\pm$&0.034 &  -1.780$\pm$0.090 \\
2016-03-28 & $\theta^1$ Ori B1-B4 & H2 &  12.260$\pm$&0.033 &  -1.780$\pm$0.080 \\
2016-03-28 & $\theta^1$ Ori B1-B4 & H3 &  12.259$\pm$&0.035 &  -1.790$\pm$0.090 \\
2016-03-30 & NGC3603 & H2 &  12.245$\pm$&0.017 &  -1.756$\pm$0.061 \\
2016-03-30 & NGC3603 & H3 &  12.240$\pm$&0.015 &  -1.767$\pm$0.061 \\
2016-04-01 & NGC3603 & K1 &  12.253$\pm$&0.013 &  -1.727$\pm$0.056 \\
2016-04-01 & NGC3603 & K2 &  12.248$\pm$&0.013 &  -1.737$\pm$0.057 \\
2016-04-16 & NGC3603 & H2 &  12.245$\pm$&0.015 &  -1.742$\pm$0.059 \\
2016-04-16 & NGC3603 & H3 &  12.242$\pm$&0.015 &  -1.752$\pm$0.056 \\
2016-04-16 & NGC3603 & K1 &  12.254$\pm$&0.013 &  -1.759$\pm$0.056 \\
2016-04-16 & NGC3603 & K2 &  12.251$\pm$&0.012 &  -1.770$\pm$0.055 \\
2016-05-22 & NGC3603 & H2 &  12.251$\pm$&0.017 &  -1.675$\pm$0.080 \\
2016-05-22 & NGC3603 & H3 &  12.244$\pm$&0.017 &  -1.689$\pm$0.081 \\
2016-05-25 & NGC3603 & K1 &  12.257$\pm$&0.011 &  -1.651$\pm$0.049 \\
2016-05-25 & NGC3603 & K2 &  12.253$\pm$&0.013 &  -1.661$\pm$0.055 \\
2016-05-31 & 47Tuc & H2 &  12.247$\pm$&0.009 &  -1.805$\pm$0.047 \\
2016-05-31 & 47Tuc & H3 &  12.240$\pm$&0.010 &  -1.812$\pm$0.047 \\
2016-06-01 & NGC3603 & H2 &  12.239$\pm$&0.014 &  -1.723$\pm$0.061 \\
2016-06-01 & NGC3603 & H3 &  12.234$\pm$&0.014 &  -1.740$\pm$0.062 \\
2016-06-01 & NGC3603 & K1 &  12.247$\pm$&0.013 &  -1.743$\pm$0.061 \\
2016-06-01 & NGC3603 & K2 &  12.243$\pm$&0.012 &  -1.755$\pm$0.058 \\
2016-06-11 & 47Tuc & H2 &  12.255$\pm$&0.010 &  -1.664$\pm$0.048 \\
2016-06-11 & 47Tuc & H3 &  12.246$\pm$&0.010 &  -1.673$\pm$0.054 \\
\rev{2016-06-14} & NGC6380 & K1 &  12.241$\pm$&0.028 &  -1.747$\pm$0.073 \\
\rev{2016-06-14} & NGC6380 & K2 &  12.237$\pm$&0.032 &  -1.769$\pm$0.080 \\
2016-06-23 & 47Tuc & H2 &  12.254$\pm$&0.011 &  -1.767$\pm$0.046 \\
2016-06-23 & 47Tuc & H3 &  12.246$\pm$&0.010 &  -1.788$\pm$0.044 \\
2016-09-16 & 47Tuc & H2 &  12.254$\pm$&0.011 &  -1.762$\pm$0.048 \\
2016-09-16 & 47Tuc & H3 &  12.244$\pm$&0.010 &  -1.777$\pm$0.043 \\
2016-10-13 & 47Tuc & H2 &  12.249$\pm$&0.011 &  -1.761$\pm$0.045 \\
2016-10-13 & 47Tuc & H3 &  12.245$\pm$&0.010 &  -1.776$\pm$0.036 \\
2016-10-14 & 47Tuc & K1 &  12.265$\pm$&0.009 &  -1.746$\pm$0.040 \\
2016-10-14 & 47Tuc & K2 &  12.258$\pm$&0.010 &  -1.767$\pm$0.040 \\
2016-11-18 & 47Tuc & H2 &  12.248$\pm$&0.010 &  -1.759$\pm$0.043 \\
2016-11-18 & 47Tuc & H3 &  12.245$\pm$&0.009 &  -1.775$\pm$0.034 \\
2016-11-18 & 47Tuc & H2 &  12.234$\pm$&0.012$^{b}$ &  -1.793$\pm$0.055 \\
2016-11-18 & 47Tuc & H3 &  12.230$\pm$&0.011$^{b}$ &  -1.803$\pm$0.051 \\
2016-12-13 & 47Tuc & H2 &  12.248$\pm$&0.011 &  -1.808$\pm$0.043 \\
2016-12-13 & 47Tuc & H3 &  12.242$\pm$&0.009 &  -1.818$\pm$0.040 \\
2017-02-07 & NGC3603 & H2 &  12.249$\pm$&0.013 &  -1.712$\pm$0.058 \\
2017-02-07 & NGC3603 & H3 &  12.245$\pm$&0.012 &  -1.724$\pm$0.055 \\
2017-02-07 & NGC3603 & K1 &  12.256$\pm$&0.012 &  -1.719$\pm$0.056 \\
2017-02-07 & NGC3603 & K2 &  12.252$\pm$&0.013 &  -1.729$\pm$0.057 \\
2017-02-11 & NGC3603 & J2 &  12.246$\pm$&0.017 &  -1.759$\pm$0.063 \\
2017-02-11 & NGC3603 & J3 &  12.238$\pm$&0.016 &  -1.770$\pm$0.061 \\
2017-03-19 & NGC3603 & H2 &  12.247$\pm$&0.012 &  -1.713$\pm$0.056 \\
2017-03-19 & NGC3603 & H3 &  12.243$\pm$&0.012 &  -1.723$\pm$0.054 \\
2017-04-28 & NGC3603 & H2 &  12.240$\pm$&0.013 &  -1.768$\pm$0.058 \\
2017-04-28 & NGC3603 & H3 &  12.236$\pm$&0.012 &  -1.781$\pm$0.056 \\
2017-04-29 & NGC3603 & K1 &  12.254$\pm$&0.012 &  -1.780$\pm$0.054 \\
2017-04-29 & NGC3603 & K2 &  12.250$\pm$&0.011 &  -1.790$\pm$0.054 \\
2017-05-09 & NGC3603 & H2 &  12.240$\pm$&0.016 &  -1.782$\pm$0.064 \\
2017-05-09 & NGC3603 & H3 &  12.235$\pm$&0.015 &  -1.794$\pm$0.058 \\
2017-05-09 & NGC3603 & K1 &  12.248$\pm$&0.012 &  -1.800$\pm$0.052 \\
2017-05-09 & NGC3603 & K2 &  12.245$\pm$&0.012 &  -1.813$\pm$0.050 \\
2017-06-02 & 47Tuc & H2 &  12.249$\pm$&0.010 &  -1.808$\pm$0.040 \\
2017-06-02 & 47Tuc & H3 &  12.243$\pm$&0.009 &  -1.819$\pm$0.041 \\
2017-06-02 & 47Tuc & K1 &  12.258$\pm$&0.015 &  -1.825$\pm$0.071 \\
2017-06-02 & 47Tuc & K2 &  12.252$\pm$&0.014 &  -1.835$\pm$0.062 \\
2017-06-12 & NGC6380 & H2 &  12.236$\pm$&0.028 &  -1.739$\pm$0.092 \\
2017-06-12 & NGC6380 & H3 &  12.231$\pm$&0.029 &  -1.753$\pm$0.088 \\
2017-09-29 & 47Tuc & H2 &  12.250$\pm$&0.010 &  -1.735$\pm$0.043 \\
2017-09-29 & 47Tuc & H3 &  12.243$\pm$&0.009 &  -1.747$\pm$0.043 \\
2017-11-03 & 47Tuc & H2 &  12.250$\pm$&0.014 &  -1.745$\pm$0.053 \\
2017-11-03 & 47Tuc & H3 &  12.242$\pm$&0.013 &  -1.756$\pm$0.051 \\
2017-11-06 & 47Tuc & K1 &  12.260$\pm$&0.011 &  -1.763$\pm$0.051 \\
2017-11-06 & 47Tuc & K2 &  12.255$\pm$&0.012 &  -1.778$\pm$0.056 \\
2017-11-30 & 47Tuc & H2 &  12.244$\pm$&0.018 &  -1.780$\pm$0.070 \\
2017-11-30 & 47Tuc & H3 &  12.243$\pm$&0.013 &  -1.790$\pm$0.056 \\
2017-11-30 & 47Tuc & K1 &  12.259$\pm$&0.009 &  -1.784$\pm$0.039 \\
2017-11-30 & 47Tuc & K2 &  12.253$\pm$&0.012 &  -1.786$\pm$0.054 \\
2018-01-03 & 47Tuc & H2 &  12.246$\pm$&0.013 &  -1.798$\pm$0.060 \\
2018-01-03 & 47Tuc & H3 &  12.240$\pm$&0.014 &  -1.807$\pm$0.059 \\
2018-01-04 & 47Tuc & K1 &  12.252$\pm$&0.010 &  -1.826$\pm$0.045 \\
2018-01-04 & 47Tuc & K2 &  12.250$\pm$&0.010 &  -1.840$\pm$0.038 \\
2018-01-25 & NGC3603 & H2 &  12.253$\pm$&0.013 &  -1.764$\pm$0.059 \\
2018-01-25 & NGC3603 & H3 &  12.247$\pm$&0.013 &  -1.777$\pm$0.060 \\
2018-01-25 & NGC3603 & K1 &  12.262$\pm$&0.014 &  -1.788$\pm$0.062 \\
2018-01-25 & NGC3603 & K2 &  12.258$\pm$&0.015 &  -1.799$\pm$0.067 \\
2018-02-24 & NGC3603 & H2 &  12.242$\pm$&0.014 &  -1.754$\pm$0.059 \\
2018-02-24 & NGC3603 & H3 &  12.239$\pm$&0.013 &  -1.765$\pm$0.055 \\
2018-02-24 & NGC3603 & K1 &  12.251$\pm$&0.013 &  -1.769$\pm$0.057 \\
2018-02-24 & NGC3603 & K2 &  12.247$\pm$&0.012 &  -1.782$\pm$0.055 \\
2018-03-28 & NGC3603 & H2 &  12.249$\pm$&0.016 &  -1.734$\pm$0.066 \\
2018-03-28 & NGC3603 & H3 &  12.246$\pm$&0.015 &  -1.745$\pm$0.063 \\
2018-04-10 & NGC3603 & H2 &  12.242$\pm$&0.017 &  -1.751$\pm$0.070 \\
2018-04-10 & NGC3603 & H3 &  12.238$\pm$&0.015 &  -1.763$\pm$0.066 \\
2018-04-10 & NGC3603 & K1 &  12.250$\pm$&0.013 &  -1.743$\pm$0.058 \\
2018-04-10 & NGC3603 & K2 &  12.246$\pm$&0.012 &  -1.754$\pm$0.056 \\
2018-05-05 & NGC3603 & H2 &  12.246$\pm$&0.016 &  -1.761$\pm$0.063 \\
2018-05-05 & NGC3603 & H3 &  12.242$\pm$&0.015 &  -1.773$\pm$0.059 \\
2018-06-18 & NGC3603 & H2 &  12.243$\pm$&0.017 &  -1.792$\pm$0.069 \\
2018-06-18 & NGC3603 & H3 &  12.239$\pm$&0.016 &  -1.801$\pm$0.065 \\
2018-09-16 & 47Tuc & H2 &  12.241$\pm$&0.017 &  -1.796$\pm$0.068 \\
2018-09-16 & 47Tuc & H3 &  12.233$\pm$&0.013 &  -1.792$\pm$0.063 \\
\rev{2018-09-19} & 47Tuc & H2 &  12.244$\pm$&0.016 &  -1.769$\pm$0.071 \\
\rev{2018-09-19} & 47Tuc & H3 &  12.239$\pm$&0.015 &  -1.787$\pm$0.066 \\
2018-10-17 & 47Tuc & H2 &  12.250$\pm$&0.009 &  -1.804$\pm$0.043 \\
2018-10-17 & 47Tuc & H3 &  12.244$\pm$&0.010 &  -1.805$\pm$0.045 \\
\rev{2018-12-15} & 47Tuc & H2 &  12.251$\pm$&0.010 &  -1.769$\pm$0.047 \\
\rev{2018-12-15} & 47Tuc & H3 &  12.245$\pm$&0.011 &  -1.780$\pm$0.048 \\
2019-03-05 & NGC3603 & K1 &  12.255$\pm$&0.013 &  -1.718$\pm$0.060 \\
2019-03-05 & NGC3603 & K2 &  12.250$\pm$&0.013 &  -1.731$\pm$0.058 \\
2019-03-06 & NGC3603 & H2 &  12.245$\pm$&0.014 &  -1.764$\pm$0.065 \\
2019-03-06 & NGC3603 & H3 &  12.241$\pm$&0.015 &  -1.776$\pm$0.062 \\
2019-04-13 & NGC3603 & K1 &  12.252$\pm$&0.015 &  -1.783$\pm$0.066 \\
2019-04-13 & NGC3603 & K2 &  12.248$\pm$&0.014 &  -1.791$\pm$0.063 \\
\rev{2019-05-17} & NGC3603 & H2 &  12.242$\pm$&0.015 &  -1.789$\pm$0.062 \\
\rev{2019-05-17} & NGC3603 & H3 &  12.238$\pm$&0.013 &  -1.800$\pm$0.058 \\
2019-11-27 & 47Tuc & H2 &  12.247$\pm$&0.017 &  -1.771$\pm$0.081 \\
2019-11-27 & 47Tuc & H3 &  12.245$\pm$&0.009 &  -1.812$\pm$0.044 \\
\caption{Pixel scale and North angle correction offset measured with SPHERE/IRDIS data from the SPHERE/SHINE survey. \\
$(a)$ Data obtained without coronagraph.}
\end{longtable}

\begin{longtable}{r@{}lccr@{}lc}
\multicolumn{2}{c}{UT Date} & Field & Filter & \multicolumn{2}{c}{Pixel scale} & North correction angle \\
\multicolumn{2}{c}{} & & & \multicolumn{2}{c}{(mas\,px$^{-1}$)} & ($^{\circ}$) \\
\hline
\rev{2014-}&12-02 & 47Tuc & BB\_H &  12.255$\pm$&0.010 &  -1.709$\pm$0.051 \\
2015-&04-28 & NGC6380 & BB\_H &  12.260$\pm$&0.022 &  -1.771$\pm$0.076 \\
2015-&04-28 & NGC6380 & BB\_Ks &  12.260$\pm$&0.022 &  -1.809$\pm$0.062 \\
2015-&04-28 & NGC6380 & H2 &  12.242$\pm$&0.027 &  -1.831$\pm$0.068 \\
2015-&04-28 & NGC6380 & H3 &  12.238$\pm$&0.027 &  -1.847$\pm$0.062 \\
2015-&04-28 & NGC6380 & K1 &  12.263$\pm$&0.030 &  -1.732$\pm$0.065 \\
2015-&04-28 & NGC6380 & K2 &  12.263$\pm$&0.031 &  -1.760$\pm$0.085 \\
\rev{2015-}&04-29 & NGC6380 & K1 & 12.234$\pm$&0.003 & -1.737$\pm$0.130 \\
\rev{2015-}&04-29 & NGC6380 & K2 & 12.231$\pm$&0.006 & -1.768$\pm$0.121 \\
\rev{2015-}&05-29 & NGC6380 & BB\_J &  12.256$\pm$&0.027 &  -1.701$\pm$0.056 \\
\rev{2015-}&05-29 & NGC6380 & BB\_H &  12.228$\pm$&0.021 &  -1.717$\pm$0.063 \\
\rev{2015-}&05-29 & NGC6380 & BB\_Ks &  12.243$\pm$&0.020 &  -1.716$\pm$0.057 \\
\rev{2015-}&05-29 & NGC6380 & H2 &  12.223$\pm$&0.021 &  -1.724$\pm$0.058 \\
\rev{2015-}&05-29 & NGC6380 & H3 &  12.223$\pm$&0.022 &  -1.731$\pm$0.056 \\
\rev{2015-}&05-30 & NGC6380 & H2 &  12.223$\pm$&0.021 &  -1.777$\pm$0.064 \\
\rev{2015-}&05-30 & NGC6380 & H3 &  12.219$\pm$&0.020 &  -1.785$\pm$0.061 \\
\rev{2015-}&06-02 & NGC6380 & K1 &  12.225$\pm$&0.025 &  -1.848$\pm$0.074 \\
\rev{2015-}&06-02 & NGC6380 & K2 &  12.222$\pm$&0.026 &  -1.857$\pm$0.076 \\
\rev{2015-}&06-28 & NGC6380 & H2 &  12.233$\pm$&0.021 &  -1.757$\pm$0.057 \\
\rev{2015-}&06-28 & NGC6380 & H3 &  12.231$\pm$&0.022 &  -1.767$\pm$0.053 \\
\rev{2015-}&07-09 & NGC6380 & BB\_J  & 12.231$\pm$&0.028 & -1.708$\pm$0.069 \\
\rev{2015-}&07-09 & NGC6380 & BB\_H &  12.224$\pm$&0.024 &  -1.768$\pm$0.071 \\
\rev{2015-}&07-09 & NGC6380 & BB\_Ks & 12.236$\pm$&0.023 &  -1.732$\pm$0.058 \\
\rev{2015-}&07-16 & NGC6380 & K1 &  12.230$\pm$&0.026 &  -1.783$\pm$0.072 \\
\rev{2015-}&07-16 & NGC6380 & K2 &  12.225$\pm$&0.027 &  -1.797$\pm$0.075 \\
\rev{2015-}&07-31 & NGC6380 & BB\_H &  12.226$\pm$&0.027 &  -1.752$\pm$0.083 \\
\rev{2015-}&07-31 & NGC6380 & BB\_Ks &  12.236$\pm$&0.025 &  -1.750$\pm$0.061 \\
\rev{2015-}&08-23 & NGC6380 & H2 & 12.228$\pm$&0.047 & -1.705$\pm$0.075 \\
\rev{2015-}&08-23 & NGC6380 & H3 & 12.221$\pm$&0.045 & -1.705$\pm$0.050 \\
2015-&08-24$^{a}$ & 47Tuc & H2 &  12.247$\pm$&0.014$^{b}$ &  -1.663$\pm$0.063 \\
2015-&08-24$^{a}$ & 47Tuc & H3 &  12.240$\pm$&0.011$^{b}$ &  -1.676$\pm$0.047 \\
\rev{2015-}&09-05 &  NGC6380 & BB\_J & 12.243$\pm$&0.030 & -1.849$\pm$0.027 \\
\rev{2015-}&09-05 & NGC6380 & BB\_H & 12.237$\pm$&0.022 & -1.809$\pm$0.073 \\
\rev{2015-}&09-05 &  NGC6380 & BB\_Ks & 12.252$\pm$&0.022 & -1.810$\pm$0.061 \\
\rev{2015-}&09-05 &  NGC6380 & K1 & 12.233$\pm$&0.025 & -1.813$\pm$0.081 \\
\rev{2015-}&09-05 &  NGC6380 & K2 & 12.232$\pm$&0.023 & -1.829$\pm$0.084 \\
2015-&10-03$^{a}$ & 47Tuc & H2 &  12.242$\pm$&0.010$^{b}$ &  -1.797$\pm$0.047 \\
2015-&10-03$^{a}$ & 47Tuc & H3 &  12.232$\pm$&0.010$^{b}$ &  -1.815$\pm$0.049 \\
2015-&10-06$^{a}$ & 47Tuc & H2 &  12.248$\pm$&0.012$^{b}$ &  -1.782$\pm$0.049 \\
2015-&10-06$^{a}$ & 47Tuc & H3 &  12.241$\pm$&0.012$^{b}$ &  -1.794$\pm$0.053 \\
2015-&10-29 & 47Tuc & BB\_H &  12.247$\pm$&0.009 &  -1.818$\pm$0.039 \\
2015-&10-29 & 47Tuc & H2 &  12.248$\pm$&0.017 &  -1.780$\pm$0.076 \\
2015-&10-29 & 47Tuc & H3 &  12.242$\pm$&0.010 &  -1.837$\pm$0.056 \\
2015-&11-14 & 47Tuc & H2 &  12.235$\pm$&0.011 &  -1.806$\pm$0.052 \\
2015-&11-14 & 47Tuc & H3 &  12.227$\pm$&0.012 &  -1.779$\pm$0.048 \\
2015-&11-14 & 47Tuc & K1 &  12.244$\pm$&0.010 &  -1.790$\pm$0.054 \\
2015-&11-14 & 47Tuc & K2 &  12.241$\pm$&0.010 &  -1.790$\pm$0.054 \\
2015-&11-23$^{a}$ & 47Tuc & H2 &  12.242$\pm$&0.013$^{b}$ &  -1.758$\pm$0.055 \\
2015-&11-23$^{a}$ & 47Tuc & H3 &  12.234$\pm$&0.012$^{b}$ &  -1.763$\pm$0.053 \\
\rev{2015-}&12-31 & $\theta^1$ Ori B1-B4 & H2 &  12.244$\pm$&0.030 &  -1.761$\pm$0.062 \\
\rev{2015-}&12-31 & $\theta^1$ Ori B1-B4 & H3 &  12.245$\pm$&0.030 &  -1.794$\pm$0.062 \\
\rev{2015-}&12-31 & $\theta^1$ Ori B1-B4 & K1 &  12.263$\pm$&0.030 &  -1.830$\pm$0.062 \\
\rev{2015-}&12-31 & $\theta^1$ Ori B1-B4 & K2 &  12.258$\pm$&0.030 &  -1.851$\pm$0.062 \\
\rev{2016-}&01-23 & $\theta^1$ Ori B1-B4 & H2 &  12.253$\pm$&0.030 &  -1.850$\pm$0.060 \\
\rev{2016-}&01-23 & $\theta^1$ Ori B1-B4 & H3 &  12.251$\pm$&0.030 &  -1.870$\pm$0.060 \\
\rev{2016-}&01-23 & $\theta^1$ Ori B1-B4 & K1 &  12.265$\pm$&0.030 &  -1.854$\pm$0.062 \\
\rev{2016-}&01-23 & $\theta^1$ Ori B1-B4 & K2 &  12.260$\pm$&0.030 &  -1.877$\pm$0.062 \\
\rev{2016-}&02-07 & $\theta^1$ Ori B1-B4 & BB\_H &  12.255$\pm$&0.030 &  -1.722$\pm$0.062 \\
\rev{2016-}&02-07 & $\theta^1$ Ori B1-B4 & H2 &  12.257$\pm$&0.030 &  -1.790$\pm$0.060 \\
\rev{2016-}&02-07 & $\theta^1$ Ori B1-B4 & H3 &  12.250$\pm$&0.030 &  -1.800$\pm$0.060 \\
\rev{2016-}&02-15 & $\theta^1$ Ori B1-B4 & K1 &  12.265$\pm$&0.030 &  -1.753$\pm$0.087 \\
\rev{2016-}&02-15 & $\theta^1$ Ori B1-B4 & K2 &  12.265$\pm$&0.031 &  -1.758$\pm$0.107 \\
\rev{2016-}&03-02 & $\theta^1$ Ori B1-B4 & H2 &  12.244$\pm$&0.030 &  -1.791$\pm$0.062 \\
\rev{2016-}&03-02 & $\theta^1$ Ori B1-B4 & H3 &  12.245$\pm$&0.030 &  -1.821$\pm$0.062 \\
\rev{2016-}&03-06 & NGC6380 & H2 &  12.235$\pm$&0.024 &  -1.751$\pm$0.064 \\
\rev{2016-}&03-06 & NGC6380 & H3 &  12.231$\pm$&0.024 &  -1.762$\pm$0.060 \\
\rev{2016-}&03-06 & NGC6380 & K1 &  12.234$\pm$&0.026 &  -1.766$\pm$0.083 \\
\rev{2016-}&03-06 & NGC6380 & K2 &  12.230$\pm$&0.027 &  -1.784$\pm$0.082 \\
\rev{2016-}&03-07 & $\theta^1$ Ori B1-B4 & H2 &  12.244$\pm$&0.030 &  -1.836$\pm$0.062 \\
\rev{2016-}&03-07 & $\theta^1$ Ori B1-B4 & H3 &  12.243$\pm$&0.030 &  -1.864$\pm$0.062 \\
\rev{2016-}&03-07 & $\theta^1$ Ori B1-B4 & K1 &  12.249$\pm$&0.030 &  -1.792$\pm$0.062 \\
\rev{2016-}&03-07 & $\theta^1$ Ori B1-B4 & K2 &  12.244$\pm$&0.030 &  -1.814$\pm$0.062 \\
\rev{2016-}&03-08 & $\theta^1$ Ori B1-B4 & H2 &  12.232$\pm$&0.030 &  -1.785$\pm$0.062 \\
\rev{2016-}&03-08 & $\theta^1$ Ori B1-B4 & H3 &  12.232$\pm$&0.030 &  -1.817$\pm$0.062 \\
\rev{2016-}&04-01 & $\theta^1$ Ori B1-B4 & K1 &  12.268$\pm$&0.030 &  -1.797$\pm$0.084 \\
\rev{2016-}&04-01 & $\theta^1$ Ori B1-B4 & K2 &  12.267$\pm$&0.030 &  -1.799$\pm$0.097 \\
\rev{2016-}&04-02 & $\theta^1$ Ori B1-B4 & BB\_J &  12.265$\pm$&0.030 &  -1.758$\pm$0.062 \\
\rev{2016-}&04-02 & $\theta^1$ Ori B1-B4 & BB\_H &  12.269$\pm$&0.031 &  -1.776$\pm$0.087 \\
\rev{2016-}&04-02 & $\theta^1$ Ori B1-B4 & BB\_Ks &  12.261$\pm$&0.031 &  -1.753$\pm$0.064 \\
\rev{2016-}&04-02 & $\theta^1$ Ori B1-B4 & H2 &  12.244$\pm$&0.030 &  -1.795$\pm$0.062 \\
\rev{2016-}&04-02 & $\theta^1$ Ori B1-B4 & H3 &  12.246$\pm$&0.030 &  -1.832$\pm$0.062 \\
\rev{2016-}&05-04 & NGC6380 & BB\_H &  12.238$\pm$&0.024 &  -1.720$\pm$0.075 \\
\rev{2016-}&05-04 & NGC6380 & BB\_Ks &  12.249$\pm$&0.024 &  -1.705$\pm$0.060 \\
\rev{2016-}&05-04 & NGC6380 & K1 &  12.230$\pm$&0.029 &  -1.721$\pm$0.068 \\
\rev{2016-}&05-04 & NGC6380 & K2 &  12.225$\pm$&0.027 &  -1.736$\pm$0.069 \\
\rev{2016-}&06-04 & NGC6380 & BB\_H &  12.227$\pm$&0.023 &  -1.772$\pm$0.075 \\
\rev{2016-}&06-04 & NGC6380 & BB\_Ks &  12.234$\pm$&0.020 &  -1.768$\pm$0.058 \\
\rev{2016-}&06-04 & NGC6380 & H2 &  12.221$\pm$&0.021 &  -1.763$\pm$0.064 \\
\rev{2016-}&06-04 & NGC6380 & H3 &  12.218$\pm$&0.023 &  -1.781$\pm$0.065 \\
\rev{2016-}&06-04 & NGC6380 & K1 &  12.226$\pm$&0.023 &  -1.807$\pm$0.081 \\
\rev{2016-}&06-04 & NGC6380 & K2 &  12.223$\pm$&0.023 &  -1.816$\pm$0.078 \\
\rev{2016-}&07-07 & NGC6380 & BB\_H &  12.242$\pm$&0.023 &  -1.779$\pm$0.075 \\
\rev{2016-}&07-07 & NGC6380 & BB\_Ks &  12.249$\pm$&0.021 &  -1.771$\pm$0.082 \\
2016-&07-18 & 47Tuc & H2 &  12.230$\pm$&0.012 &  -1.791$\pm$0.055 \\
2016-&07-18 & 47Tuc & H3 &  12.229$\pm$&0.009 &  -1.792$\pm$0.048 \\
2016-&07-18$^{a}$ & 47Tuc & H2 &  12.221$\pm$&0.011 &  -1.791$\pm$0.052 \\
2016-&07-18$^{a}$ & 47Tuc & H3 &  12.220$\pm$&0.013 &  -1.801$\pm$0.054 \\
2016-&08-18 & 47Tuc & K1 &  12.253$\pm$&0.007 &  -1.758$\pm$0.034 \\
2016-&08-18 & 47Tuc & K2 &  12.246$\pm$&0.005 &  -1.771$\pm$0.031 \\
2016-&09-01 & 47Tuc & H2 &  12.237$\pm$&0.011 &  -1.758$\pm$0.053 \\
2016-&09-01 & 47Tuc & H3 &  12.230$\pm$&0.009 &  -1.780$\pm$0.050 \\
2016-&09-01 & 47Tuc & H2 &  12.253$\pm$&0.013$^{b}$ &  -1.773$\pm$0.058 \\ 
2016-&09-01 & 47Tuc & H3 &  12.246$\pm$&0.012$^{b}$ &  -1.783$\pm$0.054 \\
2016-&09-20 & 47Tuc & BB\_H &  12.244$\pm$&0.015 &  -1.767$\pm$0.060 \\
2016-&09-20 & 47Tuc & H2 &  12.244$\pm$&0.010 &  -1.788$\pm$0.064 \\
2016-&09-20 & 47Tuc & H3 &  12.235$\pm$&0.008 &  -1.790$\pm$0.046 \\
2016-&09-20 & 47Tuc & K1 &  12.262$\pm$&0.020 &  -1.772$\pm$0.085 \\
2016-&09-20 & 47Tuc & K2 &  12.260$\pm$&0.014 &  -1.812$\pm$0.046 \\
\rev{2016-}&09-20 & $\theta^1$ Ori B1-B4 & BB\_Y &  12.291$\pm$&0.030 &  -1.823$\pm$0.062 \\
\rev{2016-}&09-20 & $\theta^1$ Ori B1-B4 & BB\_J &  12.278$\pm$&0.030 &  -1.807$\pm$0.062 \\
\rev{2016-}&09-20 & $\theta^1$ Ori B1-B4 & BB\_H &  12.268$\pm$&0.031 &  -1.807$\pm$0.064 \\
\rev{2016-}&09-20 & $\theta^1$ Ori B1-B4 & BB\_Ks &  12.268$\pm$&0.034 &  -1.797$\pm$0.083 \\
\rev{2016-}&12-05 & $\theta^1$ Ori B1-B4 & BB\_Y &  12.266$\pm$&0.030 &  -1.793$\pm$0.062 \\
\rev{2016-}&12-05 & $\theta^1$ Ori B1-B4 & BB\_J &  12.262$\pm$&0.030 &  -1.789$\pm$0.062 \\
\rev{2016-}&12-05 & $\theta^1$ Ori B1-B4& BB\_H &  12.255$\pm$&0.031 &  -1.797$\pm$0.062 \\
\rev{2016-}&12-05 & $\theta^1$ Ori B1-B4 & BB\_Ks &  12.261$\pm$&0.030 &  -1.774$\pm$0.065 \\
\rev{2016-}&12-15 & $\theta^1$ Ori B1-B4 & H2 &  12.253$\pm$&0.030 &  -1.805$\pm$0.062 \\
\rev{2016-}&12-15 & $\theta^1$ Ori B1-B4 & H3 &  12.250$\pm$&0.030 &  -1.843$\pm$0.062 \\
\rev{2016-}&12-15 & $\theta^1$ Ori B1-B4& K1 &  12.263$\pm$&0.031 &  -1.783$\pm$0.065 \\
\rev{2016-}&12-15 & $\theta^1$ Ori B1-B4 & K2 &  12.263$\pm$&0.030 &  -1.791$\pm$0.076 \\
2017-&02-04$^{a}$ & $\theta^1$ Ori B1-B4& H2 &  12.271$\pm$&0.030$^{b}$ &  -1.789$\pm$0.062 \\
2017-&02-04$^{a}$ & $\theta^1$ Ori B1-B4 & H3 &  12.257$\pm$&0.030$^{b}$ &  -1.823$\pm$0.062 \\
\rev{2017-}&02-16 & $\theta^1$ Ori B1-B4 & BB\_J &  12.269$\pm$&0.030 &  -1.767$\pm$0.062 \\
\rev{2017-}&02-16 & $\theta^1$ Ori B1-B4 & BB\_H &  12.261$\pm$&0.031 &  -1.787$\pm$0.066 \\
\rev{2017-}&02-16 & $\theta^1$ Ori B1-B4 & BB\_Ks &  12.262$\pm$&0.031 &  -1.765$\pm$0.077 \\
\rev{2017-}&03-14 & $\theta^1$ Ori B1-B4 & H2 &  12.255$\pm$&0.030 &  -1.767$\pm$0.062 \\
\rev{2017-}&03-14 & $\theta^1$ Ori B1-B4 & H3 &  12.255$\pm$&0.030 &  -1.796$\pm$0.062 \\
\rev{2017-}&03-14 & $\theta^1$ Ori B1-B4 & K1 &  12.266$\pm$&0.030 &  -1.778$\pm$0.062 \\
\rev{2017-}&03-14 & $\theta^1$ Ori B1-B4 & K2 &  12.262$\pm$&0.030 &  -1.799$\pm$0.062 \\
\rev{2017-}&05-15 & NGC6380 & BB\_J  & 12.241$\pm$&0.046 & -1.799$\pm$0.112 \\
\rev{2017-}&05-15 & NGC6380 & BB\_H &  12.229$\pm$&0.025 &  -1.755$\pm$0.077 \\
\rev{2017-}&05-15 & NGC6380 & BB\_Ks &  12.245$\pm$&0.023 &  -1.751$\pm$0.075 \\
\rev{2017-}&05-19 & NGC6380 & H2 &  12.220$\pm$&0.023 &  -1.740$\pm$0.078 \\
\rev{2017-}&05-19 & NGC6380 & H3 &  12.218$\pm$&0.024 &  -1.747$\pm$0.075 \\
\rev{2017-}&05-19 & NGC6380 & K1 &  12.223$\pm$&0.027 &  -1.735$\pm$0.097 \\
\rev{2017-}&05-19 & NGC6380 & K2 &  12.210$\pm$&0.027 &  -1.757$\pm$0.099 \\
2017-&08-10 & 47Tuc & BB\_H &  12.235$\pm$&0.016 &  -1.771$\pm$0.069 \\
2017-&11-03 & 47Tuc & BB\_H &  12.245$\pm$&0.011 &  -1.759$\pm$0.047 \\
2017-&11-12 & 47Tuc & H2 &  12.242$\pm$&0.013 &  -1.763$\pm$0.062 \\
2017-&11-12 & 47Tuc & H3 &  12.240$\pm$&0.013 &  -1.779$\pm$0.064 \\
\rev{2018-}&02-13 & $\theta^1$ Ori B1-B4 & BB\_J &  12.271$\pm$&0.030 &  -1.761$\pm$0.062 \\
\rev{2018-}&02-13 & $\theta^1$ Ori B1-B4 & BB\_H &  12.263$\pm$&0.032 &  -1.750$\pm$0.065 \\
\rev{2018-}&02-13 & $\theta^1$ Ori B1-B4 & BB\_Ks &  12.265$\pm$&0.030 &  -1.729$\pm$0.064 \\
\rev{2018-}&02-22 & NGC6380 & H2 &  12.236$\pm$&0.024 &  -1.752$\pm$0.077 \\
\rev{2018-}&02-22 & NGC6380 & H3 &  12.233$\pm$&0.024 &  -1.761$\pm$0.075 \\
\rev{2018-}&05-25 & NGC6380 & H2 &  12.230$\pm$&0.024 &  -1.711$\pm$0.089 \\
\rev{2018-}&05-25 & NGC6380 & H3 &  12.227$\pm$&0.025 &  -1.730$\pm$0.082 \\
\rev{2018-}&05-25 & NGC6380 & BB\_J & 12.253$\pm$&0.023 & -1.730$\pm$0.097 \\
\rev{2018-}&05-25 & NGC6380 & BB\_H &  12.233$\pm$&0.026 &  -1.735$\pm$0.081 \\\rev{2018-}&05-25 & NGC6380 & BB\_Ks &  12.248$\pm$&0.025 &  -1.726$\pm$0.087 \\
\rev{2018-}&06-02 & NGC6380 & BB\_H &  12.230$\pm$&0.028 &  -1.711$\pm$0.092 \\
\rev{2018-}&06-02 & NGC6380 & BB\_Ks & 12.242$\pm$&0.024 & -1.702$\pm$0.088 \\
\rev{2018-}&06-05 & NGC6380 & H2 &  12.230$\pm$&0.024 &  -1.731$\pm$0.080 \\
\rev{2018-}&06-05 & NGC6380 & H3 &  12.227$\pm$&0.025 &  -1.743$\pm$0.078 \\
\rev{2018-}&06-06 & NGC6380 & H2 &  12.232$\pm$&0.024 &  -1.805$\pm$0.083 \\
\rev{2018-}&06-06 & NGC6380 & H3 &  12.229$\pm$&0.025 &  -1.813$\pm$0.078 \\
2018-&08-07 & 47Tuc & BB\_H &  12.238$\pm$&0.010 &  -1.765$\pm$0.051 \\
2018-&08-07 & 47Tuc & H2 &  12.231$\pm$&0.013 &  -1.785$\pm$0.067 \\
2018-&08-07 & 47Tuc & H3 &  12.228$\pm$&0.012 &  -1.791$\pm$0.059 \\
\rev{2018-}&08-20 & NGC6380 & BB\_J & 12.238$\pm$&0.047 & -1.751$\pm$0.070 \\
\rev{2018-}&08-20 & NGC6380 & BB\_H &  12.230$\pm$&0.028 &  -1.721$\pm$0.094 \\
\rev{2018-}&08-20 & NGC6380 & BB\_Ks &  12.242$\pm$&0.025 &  -1.717$\pm$0.086 \\
2018-&10-28 & 47Tuc & BB\_H &  12.247$\pm$&0.011 &  -1.777$\pm$0.042 \\
\rev{2018-}&11-11 & 47Tuc & H2 &  12.248$\pm$&0.008 &  -1.762$\pm$0.065 \\
\rev{2018-}&11-11 & 47Tuc & H3 &  12.242$\pm$&0.008 &  -1.777$\pm$0.076 \\
\rev{2019-}&01-11 & $\theta^1$ Ori B1-B4 & BB\_Y &  12.276$\pm$&0.030 &  -1.817$\pm$0.062 \\
\rev{2019-}&01-11 & $\theta^1$ Ori B1-B4 & BB\_J &  12.262$\pm$&0.030 &  -1.802$\pm$0.062 \\
\rev{2019-}&01-11 & $\theta^1$ Ori B1-B4 & BB\_H &  12.258$\pm$&0.030 &  -1.784$\pm$0.066 \\
\rev{2019-}&01-11 & $\theta^1$ Ori B1-B4 & BB\_Ks &  12.262$\pm$&0.030 &  -1.762$\pm$0.065 \\
\rev{2019-}&02-21 & $\theta^1$ Ori B1-B4 & H2 &  12.246$\pm$&0.030 &  -1.751$\pm$0.062 \\
\rev{2019-}&02-21 & $\theta^1$ Ori B1-B4 & H3 &  12.253$\pm$&0.030 &  -1.750$\pm$0.062 \\
\rev{2019-}&02-22 & NGC6380 & H2 &  12.228$\pm$&0.028 &  -1.729$\pm$0.084 \\
\rev{2019-}&02-22 & NGC6380 & H3 &  12.227$\pm$&0.024 &  -1.739$\pm$0.082 \\
\rev{2019-}&05-10 & NGC6380 & BB\_J &  12.225$\pm$&0.024 &  -1.748$\pm$0.110 \\
\rev{2019-}&05-10 & NGC6380 & BB\_H &  12.228$\pm$&0.029 &  -1.771$\pm$0.105 \\
\rev{2019-}&05-10 & NGC6380 & BB\_Ks &  12.248$\pm$&0.025 &  -1.766$\pm$0.094 \\
\rev{2019-}&05-31 & NGC6380 & H2 &  12.226$\pm$&0.023 &  -1.831$\pm$0.090 \\
\rev{2019-}&05-31 & NGC6380 & H3 &  12.223$\pm$&0.024 &  -1.835$\pm$0.086 \\
\rev{2019-}&08-02 & NGC6380 & BB\_H &  12.229$\pm$&0.027 &  -1.809$\pm$0.085 \\
\rev{2019-}&08-02 & NGC6380 & BB\_Ks &  12.244$\pm$&0.026 &  -1.819$\pm$0.097 \\
\rev{2019-}&08-06 & NGC6380 & H2 &  12.224$\pm$&0.026 &  -1.696$\pm$0.086 \\
\rev{2019-}&08-06 & NGC6380 & H3 &  12.223$\pm$&0.027 &  -1.698$\pm$0.091 \\
2019-&10-25 & 47Tuc & H2 &  12.244$\pm$&0.009 &  -1.806$\pm$0.051 \\
2019-&10-25 & 47Tuc & H3 &  12.238$\pm$&0.010 &  -1.821$\pm$0.058 \\
\rev{2019-}&12-18 & $\theta^1$ Ori B1-B4 & BB\_J &  12.263$\pm$&0.030 &  -1.803$\pm$0.062 \\
\rev{2019-}&12-18 & $\theta^1$ Ori B1-B4 & BB\_H &  12.266$\pm$&0.030 &  -1.807$\pm$0.062 \\
\rev{2019-}&12-18 & $\theta^1$ Ori B1-B4 & BB\_Ks &  12.278$\pm$&0.033 &  -1.800$\pm$0.062 \\
\caption{Pixel scale and correction angle to the North measured with the ESO calibration data. $(a)$ Data obtained as part of the technical time program. $(b)$ Data obtained with coronagraph.}
\label{tab:astrocaleso}
\end{longtable}

\subsection*{Disclosures}
The authors have no relevant financial interests and no other potential conflicts of interest to disclose.

\subsection* {Acknowledgments}
The authors thank the ESO Paranal Staff for support in conducting the observations. A.L.M. acknowledges financial support from the European Research Council under the European Union’s Horizon 2020 research and innovation program (Grant Agreement No. 819155). \rev{A.V. acknowledges funding from the European Research Council (ERC) under the European Union's Horizon 2020 research and innovation programme (grant agreement No.~757561).} This work has made use of the SPHERE Data Center, jointly operated by OSUG/IPAG (Grenoble), PYTHEAS/LAM/CeSAM (Marseille), OCA/Lagrange (Nice), Observatoire de Paris/LESIA (Paris), and Observatoire de Lyon, also supported by a grant from Labex OSUG@2020 (Investissements d’avenir - ANR10 LABX56). SPHERE is an instrument designed and built by a consortium consisting of IPAG (Grenoble, France), MPIA (Heidelberg, Germany), LAM (Marseille, France), LESIA (Paris, France), Laboratoire Lagrange (Nice, France), INAF - Osservatorio di Padova (Italy), Observatoire de Gen\`eve (Switzerland), ETH Zurich (Switzerland), NOVA (Netherlands), ONERA (France), and ASTRON (Netherlands), in collaboration with ESO. SPHERE was funded by ESO, with additional contributions from CNRS (France), MPIA (Germany), INAF (Italy), FINES (Switzerland), and NOVA (Netherlands). SPHERE received funding from the European Commission Sixth and Seventh Framework Programs as part of the Optical Infrared Coordination Network for Astronomy (OPTICON) under grant number RII3-Ct-2004-001566 for FP6 (2004-2008), grant number 226604 for FP7 (2009-2012), and grant number 312430 for FP7 (2013-2016).

\subsection* {Data, Materials, and Code Availability} 
This work is based on observations collected at the European Organisation for Astronomical Research in the Southern Hemisphere under ESO programmes 095.C-0298, 095.C-0309, 096.C-0241, 097.C-0865, 198.C-0209, 1100.C-0481, 1104.C-0416, 60.A-9249, 60.A-9255, and 60.A-9800.


\bibliography{../../biblio}   
\bibliographystyle{spiejour}   



\vspace{1ex}
\noindent Biographies for the authors are not available.


\end{spacing}
\end{document}